\documentclass[10pt,english,twocolumn]{revtex4-1}

\usepackage[T1]{fontenc}

\usepackage[utf8]{inputenc} %latin9
\usepackage{geometry}
\usepackage{units}
\usepackage{textcomp}
\usepackage{amsmath}
\usepackage{babel}
\usepackage{graphicx}
\usepackage{xcolor}
\usepackage{braket}
\usepackage{float}

\begin{document}

\makeatother

\author{L. Balembois$^{1}$, J. Travesedo$^{1}$, L. Pallegoix$^{1}$, A. May$^{1,2}$, E. Billaud$^{1}$, M. Villiers$^{3}$, D. Estève$^{1}$, D. Vion$^{1}$, P. Bertet$^{1}$, E. Flurin$^{1*}$}

\affiliation{$^1$Universit\'e Paris-Saclay, CEA, CNRS, SPEC, 91191 Gif-sur-Yvette Cedex, France\\$^2$Alice$\&$Bob, 53 boulevard du Général Martial Valin, 75015 Paris \\$^3$Laboratoire de Physique de l’Ecole Normale Supr\'erieure, Mines Paris, Centre Automatique et Systemes, Inria, ENS-PSL, Universit\'e PSL, CNRS, Sorbonne Universit\'e, Paris, France}

\date{\today}

\email{emmanuel.flurin@cea.fr}

\title{Cyclically operated Single Microwave Photon Counter with $10^\mathrm{-22}$ $\mathrm{W/\sqrt{Hz}}$ sensitivity.}

\begin{abstract}

Single photon detection played an important role in the development of quantum optics. Its implementation in the microwave domain is challenging because the photon energy is 5 orders of magnitude smaller. In recent years, significant progress has been made in developing single microwave photon detectors (SMPDs) based on superconducting quantum bits or bolometers. In this paper we present a practical SMPD based on the irreversible transfer of an incoming photon to the excited state of a transmon qubit by a four-wave mixing process. This device achieves a detection efficiency $\eta = 0.43$ and an operational dark count rate $\alpha = 85$ $\mathrm{s^{-1}}$, mainly due to the out-of-equilibrium microwave photons in the input line. The corresponding power sensitivity is $\mathcal{S} = 10^{-22}$ $\mathrm{W/\sqrt{Hz}}$, one order of magnitude lower than the state of the art. The detector operates continuously over hour timescales with a duty cycle $\eta_\mathrm{D}=0.84$, and offers frequency tunability of at least 50 MHz around 7 GHz.

\end{abstract}

\maketitle

%Single photon detection in the optical domain is a key enabling technology for many applications, ranging from fluorescence microscopy \cite{orrit_single_1990,klar_fluorescence_2000,betzig_imaging_2006,bruschini_single-photon_2019} to measurement-based quantum computing \cite{hadfield_single-photon_2009}. 

%In the microwave domain, single-photon detectors (SMPD) operating at millikelvin temperatures have only recently started to be developed, due to the 5 orders of magnitude difference in photon energy. Designs based either on superconducting quantum bits or bolometers have been proposed~\cite{romero_microwave_2009,helmer_quantum_2009,sathyamoorthy_quantum_2014,kyriienko_continuous-wave_2016,sathyamoorthy_detecting_2016,gu_microwave_2017,wong_quantum_2017,royer_itinerant_2018,grimsmo_quantum_2020} and implemented~\cite{chen_microwave_2011,koshino_implementation_2013,narla_robust_2016,inomata_single_2016,besse_single-shot_2018,kono_quantum_2018,lee_graphene-based_2020}. Besides itinerant microwave photon detectors, other experiments have demonstrated high sensitivity detection of individual microwave photons in a high-Q cavity \cite{Schuster2007-zt, Gleyzes2007-xp, dixit_searching_2021}. 

Single photon detection is a mature technique in the optical domain. Its applications are numerous, ranging from fluorescence microscopy \cite{orrit_single_1990,klar_fluorescence_2000,betzig_imaging_2006,bruschini_single-photon_2019} to measurement-based quantum computing \cite{hadfield_single-photon_2009}. At microwave frequencies ($5-10$\,GHz), single-photon detection is more challenging due to the five orders of magnitude photon energy, requiring in particular millikelvin temperatures to minimize the number of thermal photons per mode. Nevertheless, a range of applications make the development of such devices relevant. Single Microwave Photon Detectors (SMPDs) have been proposed for detecting weak incoherent emitters at microwave frequencies, such as electron spins in solids~\cite{albertinale_detecting_2021,billaud_microwave_2022, wang_single_2023}  or hypothetical dark matter candidate particles~\cite{ lamoreaux_analysis_2013, dixit_searching_2021}. SMPDs may also be useful for primary thermometry~\cite{scigliuzzo_primary_2020} and for quantum illumination protocols~\cite{assouly_quantum_2023}. Finally, SMPDs will enable the implementation of several quantum information processing protocols~\cite{raussendorf_measurement-based_2003,briegel_measurement-based_2009,bartolucci_fusion-based_2021}, for instance for the heralded entanglement of superconducting qubits at a distance~\cite{narla_robust_2016}, the development of new qubit readout schemes \cite{opremcak_measurement_2018}, or the robust generation of quantum states \cite{besse_parity_2020}. 

%The advent of these SMPDs has already enabled new classes of protocols for quantum sensing such as the microwave fluorescence detection of small electronic spin ensemble \cite{albertinale_detecting_2021,billaud_microwave_2022} and dark-matter search based on haloscopes \cite{lamoreaux_analysis_2013, dixit_searching_2021}, but also for quantum computing\cite{raussendorf_measurement-based_2003,briegel_measurement-based_2009,bartolucci_fusion-based_2021} with new superconducting qubit readout \cite{opremcak_measurement_2018} or the robust generation of quantum states \cite{besse_parity_2020}. 

% SMPDs 

SMPD designs based either on superconducting quantum bits or bolometers have been proposed~\cite{romero_microwave_2009,helmer_quantum_2009,sathyamoorthy_quantum_2014,kyriienko_continuous-wave_2016,sathyamoorthy_detecting_2016,gu_microwave_2017,wong_quantum_2017,royer_itinerant_2018,grimsmo_quantum_2020} and implemented~\cite{chen_microwave_2011,koshino_implementation_2013,narla_robust_2016,inomata_single_2016,besse_single-shot_2018,kono_quantum_2018,lee_graphene-based_2020}. Besides itinerant microwave photon detectors, other experiments have demonstrated high sensitivity detection of individual microwave photons in a high-Q cavity \cite{Schuster2007-zt, Gleyzes2007-xp, dixit_searching_2021}. 

The detection sensitivity, as well as the fidelity of the envisioned quantum protocols, depend crucially on the SMPD characteristics. Two figures of merit especially matter: the dark count rate $\alpha$ defined as the number of false positive detection per unit of time, and the operational efficiency $\eta$ defined as the ratio of counts over incoming photons. Combining these two metrics, one can determine the power sensitivity $\mathcal{S}$ of the detector as the noise equivalent power (NEP) for an integration time of 1 s (see Appendix C):

\begin{equation}
    \mathcal{S}=\dfrac{\hbar \omega \sqrt{\alpha}}{\eta}.
    \label{eq:NEP}
\end{equation}

Currently, the detectors based on superconducting qubit \cite{ inomata_single_2016,besse_single-shot_2018,kono_quantum_2018} show a dark count rate $\alpha \sim 10^\mathrm{4}-10^\mathrm{5} \mathrm{s^{-1}}$, for an efficiency $\eta \sim 0.5-0.7$ over a bandwidth of $\sim 10-20$ MHz, resulting in a sensitivity $\mathcal{S} \sim 2-9\cdot 10^\mathrm{-21}$ $\mathrm{W/\sqrt{Hz}}$ at 7 GHz. On the other hand, the most advanced bolometric detector based on graphene \cite{lee_graphene-based_2020} reaches a sensitivity $\mathcal{S} = 7\cdot 10^\mathrm{-19}$ $\mathrm{W/\sqrt{Hz}}$ at 7.9 GHz, when operated at 190 mK. The bandwidth of this device varies between 861 MHz and 599 MHz depending on the operating parameters. 

This article presents a SMPD based on a superconducting qubit and a four-wave mixing process \cite{lescanne_irreversible_2020}. It detects itinerant photons, regardless of their waveform, in a $\sim$ 1 MHz bandwidth around a frequency tunable from 7.005 GHz to, at least, 6.955 GHz (see Appendix B) and operates by cycles of $\sim 12$ $\mathrm{\mu s}$ duration, which can be repeated continuously over several hours, days, or even months. Here we demonstrate, at the specific frequency of 6.979 GHz, a dark count rate $\alpha= 85$ $\mathrm{s^{-1}}$ for an operational efficiency $\eta=0.43$ leading to a power sensitivity $\mathcal{S} = 10^{-22}$ $\mathrm{W/\sqrt{Hz}}$, more than an order of magnitude lower than the state of the art. This new sensitivity has opened up new detection possibilities. In particular, a device (called SMPD2 in the following, Appendix I) very similar to the one discussed in this work has recently enabled the detection of individual electron spins in solids, through their microwave fluorescence \cite{wang_single_2023}.

%This paper presents a SMPD based on a superconducting qubit and a four-wave mixing process \cite{lescanne_irreversible_2020}. It detects itinerant photons, regardless of their waveform, in a $\sim$ 1 MHz bandwidth around a frequency tunable from 7.005 GHz to, at least, 6.955 GHz (see Supplementary Material \cite{suppmat_2023}) and operates by cycles of $\sim 12$ $\mathrm{\mu s}$ duration, which can be repeated continuously over several hours, days, or even months. Here we demonstrate, at the specific frequency of 6.979 GHz, a dark count rate $\alpha= 85$ $\mathrm{s^{-1}}$ for an operational efficiency $\eta=0.43$ leading to a power sensitivity $\mathcal{S} = 10^{-22}$ $\mathrm{W/\sqrt{Hz}}$, more than an order of magnitude lower than the state of the art. This new sensitivity has opened up new detection possibilities, such as the single-spin detection experiment \cite{wang_single_2023}, realised with a twin device (SMPD2) to the one presented in this article, SMPD1 (see Supplementary Material \cite{suppmat_2023}).

\section{Working principle}

\begin{figure}[t!]
  \includegraphics[width=\columnwidth]{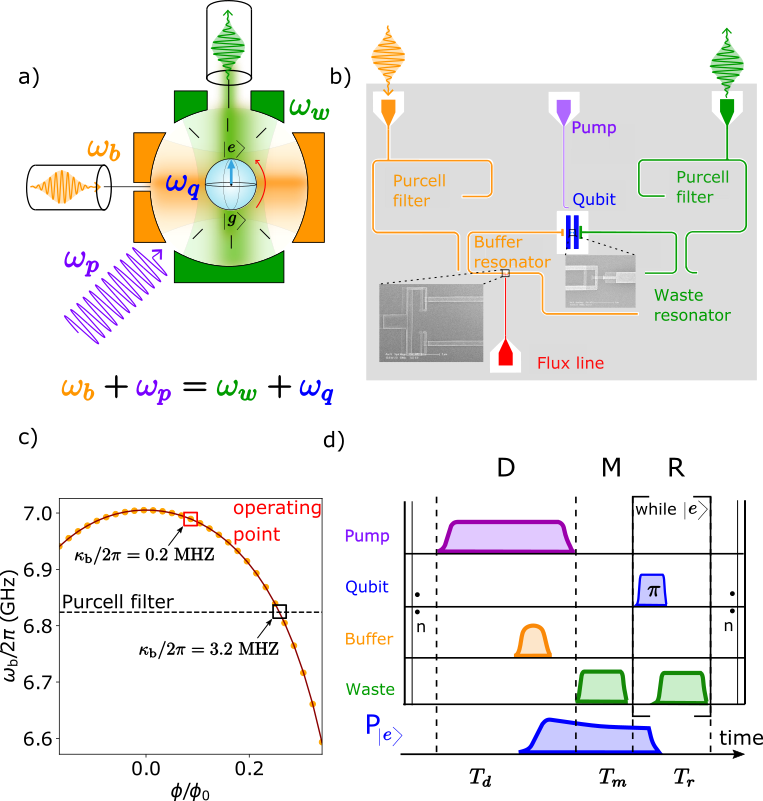}
  \caption{\label{fig1} 
  a) Principle of the photon detector. Two cavities, the buffer (orange) and the waste (green), are coupled to a transmon qubit whose non-linearity allows the modes to be mixed. A pump tone (purple) triggers a four-wave mixing process, converting an incoming buffer photon into a long-lived qubit excitation and a waste photon quickly dissipated into the environment, making the reversal process impossible.  b) Schematic of the SMPD chip. The transmon qubit (blue) at frequency $\omega_\mathrm{q}/2\pi = 6.184$ GHz is capacitively coupled to two CPW resonators: the buffer (characteristic see c.) and the waste ($\omega_\mathrm{w}/2\pi = 7.704$ GHz, $\kappa_\mathrm{w}/2\pi = 1.8$ MHz). Two Purcell filters are added to protect the qubit from radiative relaxation. The tunability of the detector is ensured by inserting a SQUID, driven by a flux line (red), in the buffer resonator. c) Evolution of the buffer frequency with the respect to the magnetic flux through the SQUID. Orange points are data, solid red line is a fit, and dashed black line represents the buffer Purcell filter frequency. Due to the frequency detuning between the buffer and its filter, the buffer bandwidth $\kappa_\mathrm{b}$ varies with its frequency. Red square represents the operating point. d) Cyclic operation of the SMPD consisting in three steps repeated continuously. The detection window (D) consists in switching on the pump tone (purple) during $T_\mathrm{d} = 10$ $\mathrm{\mu s}$ to allow the conversion of the incoming photon (orange). The measurement window (M) consists in applying a readout pulse (green) on the waste resonator to measure the qubit state. The reset window (R) is a conditional loop to reinitialize the qubit in its ground state. The excited population of the qubit during the pulse sequence is represented below. The average detector blind time is $T_\mathrm{m} + T_\mathrm{r} = 1.9$ $\mathrm{\mu s}$.     
  }
\end{figure}

This device builds upon the superconducting circuit proposed and demonstrated in \cite{lescanne_irreversible_2020} and \cite{albertinale_detecting_2021}. The working principle is based on the irreversible transfer of an incoming photon to an excitation of a transmon qubit. The detector "clicks" when the qubit is detected in its excited state using dispersive readout through a capacitively coupled resonator.

This irreversible transfer is achieved by a four-wave mixing process, directly provided by the transmon qubit Hamiltonian. The incoming photon impinging on an input resonator with frequency $\omega_\mathrm{b}$ (called "buffer" mode, orange in Fig.~\ref{fig1}a) combines with a pump tone at frequency $\omega_\mathrm{p}$ and is converted into an excitation in the transmon qubit mode at frequency $\omega_\mathrm{q}$ and an additional photon in an output resonator mode at frequency $\omega_\mathrm{w}$ (called "waste" mode, green in Fig.~\ref{fig1}a).
This four-wave mixing process is described by the Hamiltonian 

\begin{equation}
    \hat{H}_\mathrm{4WM}=\sqrt{\chi_\mathrm{b} \chi_\mathrm{w}} \left(\xi \hat{b}\hat{\sigma}^\dagger\hat{w}^\dagger+\xi^* \hat{b}^\dagger\hat{\sigma}\hat{w}\right),
    \label{eq:Hamiltonian_4Wm}
\end{equation}

where  $\hat{b},\hat{w}$ are the annihilation operators corresponding to the buffer mode and waste mode, $\hat{\sigma}$ is the lowering operator corresponding to the qubit, $\xi$ is the pump amplitude in the qubit mode, and $\chi_\mathrm{b}$, $\chi_\mathrm{w}$ are the dispersive shifts of the transmon qubit with respect to the buffer and waste modes \cite{lescanne_irreversible_2020}.
For this process to be activated, the pump frequency is tuned such that $\omega_\mathrm{p}+\omega_\mathrm{b}=\omega_\mathrm{q}+\omega_\mathrm{w}-\chi_\mathrm{w}$, to satisfy the four-wave mixing resonance condition. 

The irreversibility of the conversion is ensured by the coupling of the waste resonator to a dissipative environment. While the qubit remains excited, the photon in the waste resonator leaks out in the measurement line at the rate $\kappa_\mathrm{w}$. The reciprocal four-wave mixing process (second term in the parenthesis of Eq.\eqref{eq:Hamiltonian_4Wm}) is therefore suppressed and the qubit is left in its excited state. The detector behaves as an energy integrator, which is independent of the incoming photon waveform provided that its spectral extension remains included than the frequency linewidth of the buffer mode. 

%The energy provided by the microwave photon is stored in the transmon qubit during its typical relaxation time $T_\mathrm{1}$, which is measured at $T_\mathrm{1} \sim 37$ $\mathrm{\mu s}$ (see Fig.~\ref{fig2}d). False positives can occur in this detection scheme due to the non-zero qubit equilibrium population $p_\mathrm{eq}$, fluctuating around $p_\mathrm{eq} \sim 2-4 \cdot 10^\mathrm{-4}$ (see Fig.~\ref{fig2}c,d). The transmon qubit frequency is measured to $\omega_\mathrm{q}/2\pi = 6.184$ GHz

The four-wave mixing being a resonant process, it is intrinsically narrowband. To make it a practical detector, our device is made frequency tunable to match the photon frequency of interest by inserting a SQUID in the buffer resonator (see Fig.~\ref{fig1}b). 
The detector frequency can be tuned from $\omega_\mathrm{b}/2\pi= 7.005$  GHz over several hundred of MHz (see Fig.~\ref{fig1}c). Two band-pass Purcell filters are associated with the resonators to prevent spurious decay of the qubit into the lines \cite{korotkov_Purcell_filter_2015}. Therefore, the buffer resonator linewidth depends on its frequency detuning with respect to its Purcell filter. The bandwidth $\kappa_\mathrm{b}/2\pi= 3$ MHz is maximal for $\omega_\mathrm{b}/2\pi = 6.824$ GHz. In the following, the detector is characterized at $\omega_\mathrm{b}/2\pi = 6.979$ GHz and $\kappa_\mathrm{b}/2\pi = 0.2$ MHz. The resonance frequencies of the waste resonator $\omega_\mathrm{w}/2\pi = 7.704$ GHz and the transmon qubit $\omega_\mathrm{q}/2\pi = 6.184$ GHz are fixed. The relaxation time $T_\mathrm{1}$ of the transmon qubit is measured to $T_\mathrm{1} \sim 37$ $\mathrm{\mu s}$ (see Fig.~\ref{fig2}d) and its equilibrium population fluctuates around $p_\mathrm{eq} \sim 2-4 \cdot 10^\mathrm{-4}$  (see Fig.~\ref{fig2}c,d), close to the lowest reported \cite{serniak_hot_2018,serniak_direct_2019,Jin_Thermal_2015,connolly_coexistence_2023}. For the particular value $\omega_\mathrm{b}/2\pi = 6.935$ GHz, the pump tone set by the four-wave mixing resonance condition becomes resonant with the buffer resonator. Under such conditions, the pump couples directly to the buffer, the four-wave mixing becomes degenerate, and the SMPD cannot operate. This collision between the pump and buffer frequencies is due to fabrication inaccuracies and has been corrected in the second version of the design (SMPD2 Appendix I).

The optimal pump characteristics (frequency $\omega_\mathrm{p}/2\pi$ and amplitude $\xi$) are determined experimentally by monitoring the qubit population while illuminating the buffer mode with a weak coherent signal and by sweeping the pump tone frequency and amplitude. As shown in Fig.~\ref{fig2}b, a large excited state population is found in the qubit state conditioned on the presence of the illuminating tone for a pump frequency of $\omega_\mathrm{p}/2\pi = 6.885$ GHz. This value is in good agreement with the mode frequencies taking into account the qubit Stark shift induced by the pump and the dispersive shift of the resonators.

In a restricted subspace where the buffer and the waste are never simultaneously populated, our device can be described by two cavities coupled with the strength $g_\mathrm{3} = 2|\xi|\sqrt{\chi_\mathrm{b} \chi_\mathrm{w}}$ \cite{albertinale_detecting_2021} (see Appendix D). In this framework, the maximum detection efficiency is expected when the coupling strength $g_\mathrm{3}$ matches the geometric mean of decay rates of the buffer and waste resonators such that $2|\xi|\sqrt{\chi_\mathrm{b} \chi_\mathrm{w}} = \sqrt{\kappa_\mathrm{b}\kappa_\mathrm{w}}$ \cite{lescanne_irreversible_2020}. The model also provides an explicit formula for the transfer efficiency $\eta_\mathrm{4wm}$ between a buffer photon and a qubit excitation: 

\begin{equation}
    \eta_\mathrm{4wm}=\dfrac{4C}{(1+C)^2},
    \label{eq:eta_d}
\end{equation}

where $C=4|\xi|^2\frac{\chi_\mathrm{b} \chi_\mathrm{w}}{\kappa_\mathrm{b} \kappa_\mathrm{w}}$ is the cooperativity associated with the four-wave mixing. Unit transfer efficiency is reached for $C=1$. Taking into account resonator losses, we expect a maximum transfer efficiency of $\eta_\mathrm{4wm} = 0.86$ (see Appendix E). To determine the pump amplitude corresponding to the optimal cooperativity, we operate the four-wave mixing for various pump amplitude by sending photons on the buffer resonator. The resulting qubit excited population, plotted in Fig.~\ref{fig2}a, is in good agreement with the theoretical two-coupled cavities model. We have chosen to operate slightly below the optimum pump amplitude to mitigate the heating effects.

In order to avoid spurious qubit heating due to the pump tone, a low pump amplitude is desirable. This is conveniently achieved if the dispersive shifts $\chi_\mathrm{b,w}$ are larger than the $\kappa_\mathrm{b,w}$. Here, the measured dispersive shifts are $\chi_\mathrm{b}/2\pi=5.2\ \mathrm{MHz}$, $\chi_\mathrm{w}/2\pi=18.8\ \mathrm{MHz}$ and the resonators linewidths  $\kappa_\mathrm{b}/2\pi=0.2\ \mathrm{MHz}$ (at the point considered) and $\kappa_\mathrm{w}/2\pi=1.8\ \mathrm{MHz}$. At unit cooperativity, the pump energy in unit of qubit excitation is $|\xi|^2= 5\times10^{-3}$. Note that the large dispersive shift between the resonators and the qubit are not detrimental for two reasons. First, the maximum number of excitation in modes during the transfer process never exceeds one, so that higher order non-linear terms do not contribute significantly to the dynamics as shown on Fig. \ref{fig2}b. Second, Purcell filters at the output of each of the resonators inhibit the spurious decay of the qubit into transmission lines.

%Under standard operating conditions, we operate slightly below the optimum pump amplitude as shown in \ref{fig2}a to further reduce the pump-induced dark count rate. Taking into account the losses of the buffer resonator with a two coupled cavities model \cite{albertinale_detecting_2021}, the transfer efficiency is $\eta_\mathrm{4wm} = 0.86$.  

%, leading to qubit relaxation time $T_1=37\ \mathrm{\mu s}$. 

\section{Cyclic operation}

\begin{figure}[t!]
  \includegraphics[width=\columnwidth]{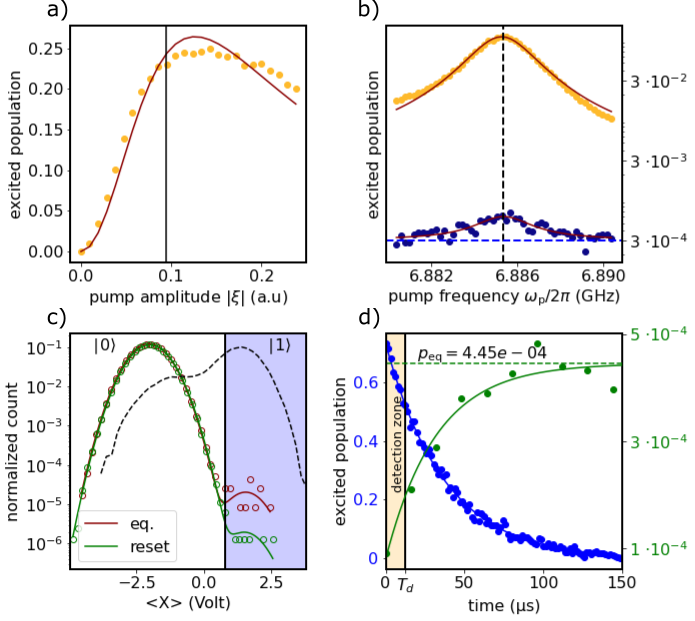}
  \caption{\label{fig2}
    a) Qubit excited population as a function of the pump amplitude $|\xi|$ applied on the qubit when a coherent tone of 360 zW (77850 photon/s) is applied on the buffer resonator. The pump frequency is adapted for each amplitude to follow the Stark shift. Orange points are data, dark red solid line represents a fit using (\ref{eq:eta_d}) and black solid line represents the chosen amplitude. b) Qubit excited population as the function of the pump frequency $\omega_\mathrm{p}/2\pi$ when no signal is sent on the buffer (dark blue) and when a coherent tone (360 zW) is applied (orange). Solid red lines represent Lorentzian fit, dashed black line is the center of the resonance. c) Qubit readout when no pulses are applied to the qubit (red) giving a qubit equilibrium population $p_\mathrm{eq}= 2\cdot10^\mathrm{-4}$ and qubit readout just after a reset sequence (green) giving a reset population $p_\mathrm{reset} =10^\mathrm{-5}$. Solid lines represent Gaussian fit. The dashed black line represents the qubit readout after a $\pi-$pulse, the vertical solid line corresponds to the chosen readout threshold.
    d) Qubit relaxation curve from the excited state (blue) and from the ground state (green). Solid lines represent exponential fits. The yellow window corresponds to the detection time $T_\mathrm{d} = 10$ $\mathrm{\mu s}$.
  }
\end{figure}

The detector is operated cyclically, with the operation cycle consisting of three subsequent steps (see Fig.~\ref{fig1}d).
The first one called "detection" (D) consists in applying to the qubit a pump pulse at frequency $\omega_\mathrm{p}$  during a detection time $T_\mathrm{d} = 10$ $\mathrm{\mu s}$ (see Appendix G for $T_\mathrm{d}$ calibration). If a photon enters into the buffer resonator, the four-wave mixing process triggers a qubit excitation and a dissipation of a photon into the waste resonator. 

In the second step of the cycle called "measurement" (M), the qubit state is dispersively readout using the waste resonator during a measurement time $T_\mathrm{m} = 0.5$ $\mathrm{\mu s}$. Note that the threshold used to discriminate the qubit ground and excited states is chosen to maximize the SMPD power sensitivity defined in Eq.\eqref{eq:NEP}. As shown in Figure~\ref{fig2}c, this threshold favors the readout fidelity of the ground state at the expense of the readout fidelity of the excited state
$p(1|e)= \eta_\mathrm{RO}=0.73$. The dark count is minimized at the expense of a moderate reduction of the efficiency.

%Note that the readout fidelity is biased in order to maximize the power sensitivity of the detector defined in Eq.\eqref{eq:NEP}. As shown in Figure~\ref{fig2}c, the readout discrimination threshold favors the readout fidelity of the ground state  $p(0|g)=1-p_\mathrm{eq}=0.9998$ at the expense of the readout fidelity of the excited state $p(1|e)= \eta_\mathrm{RO}=0.73$. These readout fidelities are chosen to minimize $\sqrt{1-p(0|g)}/p(1|e)$ which is proportional to the detector power sensitivity, the dark count is favored at the expense of a moderate reduction of the efficiency.

The third step of the cycle consists of a conditional reset (R). If the qubit is previously found in its ground state, we directly go to the next cycle. If the qubit was found in its excited state, a $\pi$-pulse is applied though the pump line and the qubit state is measured again, the procedure being repeated until the ground preparation succeeds. Owing to the high fidelity of the qubit ground state readout, we reset the qubit well below its equilibrium population $p_\mathrm{eq}$ as shown in Figure~\ref{fig2}d, the reset infidelity is as low as $p_\mathrm{reset} = 10^{-5}$. The reset step is non-deterministic, with an average reset time $T_\mathrm{r} \approx 0.5$ $\mathrm{\mu s}$. For the reset to work optimally, the Quantum Non Demolition character of the measurement must be ensured. To meet this condition, we use a Traveling Wave Parametric Amplifier (TWPA) and we tune carefully the readout pulse length and amplitude. Moreover, the readout pulse frequency is detuned with the respect to the waste resonator frequency by the dispersive shift of the qubit, such that the readout pulse enters the resonator if and only if the qubit is in its excited state. This allows to enhance the Quantum Non Demolition character of the measurement when the qubit is in its ground state. 

A waiting time of 1 $\mathrm{\mu s}$ is added at the end of the reset step to let the waste resonator return to its ground state. The average cycle time is $T_\mathrm{cycle} = 11.9$ $\mathrm{\mu s}$, which sets the duty cycle of the detector $\eta_\mathrm{D} = T_\mathrm{d}/T_\mathrm{cycle} = 0.84$. This quantity could be made arbitrary close to one by increasing the duration of the detection window. However, the qubit relaxation in a characteristic time  $T_\mathrm{1} = 37$ $\mathrm{\mu s}$ (see Fig.~\ref{fig2}d) sets an upper bound, by introducing a contribution $\eta_\mathrm{qubit} = (T_\mathrm{1}/T_\mathrm{d})(1-e^{-T_\mathrm{d} / T_\mathrm{1}})$ to the overall efficiency, which actually limits the detection step duration.

The detector is operated by continuously repeating the cycle  $\sim 8\cdot 10^4$ times per second. Its temporal resolution is determined by the detection time $T_\mathrm{d} = 10$ $\mathrm{\mu s}$, whereas its dead-time is $T_\mathrm{m}+T_\mathrm{reset} = 1.9$ $\mathrm{\mu s}$.

%The overall cycle time $T_\mathrm{cycle} = 11.9$ $\mathrm{\mu s}$ sets the time resolution of the detector. The detector is then operated by continuously repeating the cycle  $\sim$ 80000 per second. 

\section{Detection efficiency}

The operational efficiency is measured by sending a calibrated tone at the center of the SMPD line (see Fig.~\ref{fig3}d) while the cycle is repeated. The power of the microwave tone is calibrated by using the dephasing of the qubit induced by the presence of photons in the buffer cavity \cite{gambetta_qubit-photon_2006}.

Typical measurement records of the detector for various illumination powers are shown Fig.~\ref{fig3}a. The operational efficiency of the SMPD $\eta=0.43$  is obtained by measuring the ratio between the click event rate over the incoming photons rate as shown in Fig.~\ref{fig3}b. The efficiency is in good agreement with the expected one that includes four different contributions: the transfer efficiency $\eta_\mathrm{4wm}$, the qubit relaxation $\eta_\mathrm{qubit}$, the duty cycle $\eta_\mathrm{D}$ and the readout fidelity  $\eta_\mathrm{RO}$ resulting into a theoretical efficiency $\eta_\mathrm{theory} = \eta_\mathrm{4wm} \cdot \eta_\mathrm{RO} \cdot \eta_\mathrm{D} \cdot \eta_\mathrm{qubit} = 0.46$.

\section{Detection bandwidth}

\begin{figure}
  \includegraphics[width=\columnwidth]{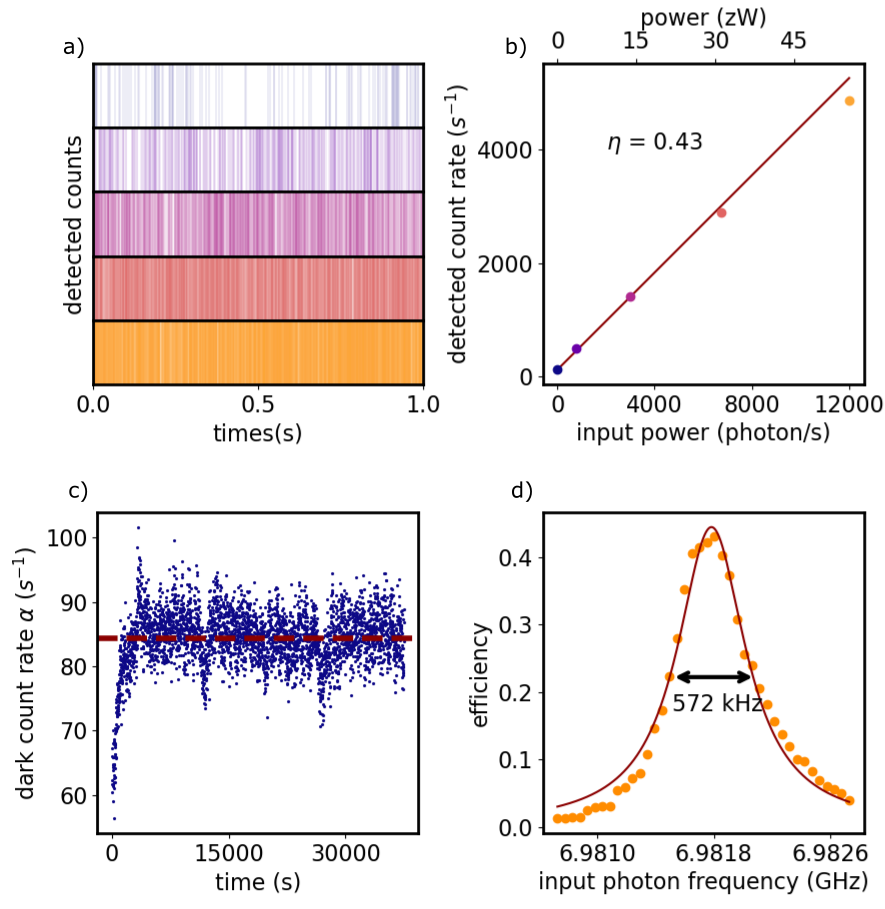}
  \caption{\label{fig3} 
  a) Time traces of the SMPD operated in cyclic mode, each vertical line represents one photon detection. The power of the coherent tone sent on the buffer resonator is progressively increased from 0W (dark blue) to 54 zW (12000 photon/s) (orange). b) Detected count rate as a function of incoming photon rate. The efficiency $\eta$ is extracted with a linear fit (solid line). The deviation to the linear behaviour is due the the detector saturation. c) dark count rate as the function of time. Each points is the average rate over $\approx 12 \mathrm{s}$ which correspond to $10^{6}$ cycles. The dashed line represent the dark count saturation $\alpha = 85$ $\mathrm{s^{-1}}$
  d). Inferred efficiency of the SMPD as a function of the frequency of the coherent tone applied on the buffer resonator. the power of the tone is still 360 zW. Solid lines represent a Lorentzian fit. The detector bandwidth defined as the FWMH is $\kappa_\mathrm{d}/2\pi = 0.57$ MHz   
  }
\end{figure}

The detector bandwidth is measured by varying the frequency of a 10 $\mathrm{\mu s}$ photon pulse sent to the buffer resonator during the detection window. The qubit excitation probability is measured and multiplied by a constant factor so that the maximum value corresponds to the overall efficiency $\eta = 0.43$. This infered efficiency is then plotted with the respect to the frequency of the input photons as shown in Fig.~\ref{fig3}d. The detector bandwidth, defined as the full width half-maximum is $\kappa_{\mathrm{d}}/2\pi = 0.57\ \mathrm{MHz}$.

From a model of two coupled cavities (see Appendix D), explicitly derived in \cite{albertinale_detecting_2021}, we can obtain an analytical expression of $\kappa_\mathrm{d}$ with respect to $\kappa_\mathrm{b}$ and $\kappa_\mathrm{w}$: 

\begin{equation}\label{eq:SMPD_BW}
\kappa_\mathrm{d}=\sqrt{2}\sqrt{\sqrt{\kappa_\mathrm{b}^2\kappa_\mathrm{w}^2+\left(\frac{\kappa_\mathrm{b}-\kappa_\mathrm{w}}{2}\right)^4}-\left(\frac{\kappa_\mathrm{b}-\kappa_\mathrm{w}}{2}\right)^2},
\end{equation}

yielding to the theoretical $\kappa_\mathrm{d,th}/2\pi = 0.43$ MHz. Here $\kappa_\mathrm{d,th} \approx 2\kappa_\mathrm{b}$ which corresponds to the limit where $\kappa_\mathrm{w} \gg \kappa_\mathrm{b}$. We attribute the discrepancy between the theoretical and the measured bandwidth  to the 100 kHz spectral broadening caused by the finite length of the excitation pulses.

%As we are in the limit where $\kappa_\mathrm{b}\ll\kappa_\mathrm{w}$, we can take a second order development of (\ref{eq:SMPD_BW}). The expected bandwidth is then given by $\frac{\kappa_{\mathrm{d}}}{2\pi}\approx 2\frac{\kappa_\mathrm{b}}{2\pi}(1+\kappa_a/\kappa_b)= 0.43\ \mathrm{MHz}$. We attribute the discrepancy between the theoretical and the measured efficiency to the 100 kHz spectral broadening caused by the finite length of the excitation pulses.

\section{dark counts}

The dark count rate is estimated by measuring the count-rate of the detector in the absence of input photons as illustrated on the top panel of Fig.~\ref{fig3}a. The dark count rate is found to be 60 $\mathrm{s^{-1}}$ for few minute of the operation. As shown in Fig.~\ref{fig3}c , when operated on hour time-scale, we observe a slight rise of the dark count rate to 85 $\mathrm{s^{-1}}$ during 1 hours, after what it remains stable within $\pm 4$ $\mathrm{s^{-1}}$ over 10 hours. The initial dark count rise is attributed to the heating of the cold stage of the refrigerator due to the continuous power delivered by the qubit pump. The sensitivity of the detector in steady state regime is then simply given by Eq. \ref{eq:NEP} and yields to a value $\mathcal{S} = 10^{-22}$ $\mathrm{W/\sqrt{Hz}}$.

\section{dark count budget}

The dark counts $\alpha$ can be decomposed in three main contributions: the thermal population of the qubit $\alpha_\mathrm{qubit}$, the heating of qubit by the pump $\alpha_\mathrm{4wm}$, and the presence of thermal photons in the input lines $\alpha_\mathrm{th}$. The resulting dark count rate is the sum of the three contributions: $\alpha = \alpha_\mathrm{qubit}+\alpha_\mathrm{4wm}+\alpha_\mathrm{th}$, each of them can be addressed individually.

The first contribution is the probability to find the qubit in its excited state in the absence of the four-wave mixing process. This depends on the qubit excitation probability after the reset $p_\mathrm{reset}$ and the relaxation rate $T_1^{-1}$ of the qubit toward its equilibrium population $p_\mathrm{eq}$ (see Fig.~\ref{fig2}c,d) such as $\alpha_{\mathrm{qubit}} = \frac{p_{\mathrm{eq}}}{T_\mathrm{1}}\eta_\mathrm{D} + \frac{p_{\mathrm{reset}}}{T_{\mathrm{cycle}}}$. We evaluate this contribution to $\alpha_{\mathrm{qubit}} = 5$ $\mathrm{s^{-1}}$ by using the parameters $p_\mathrm{eq}, T_\mathrm{1}, p_\mathrm{reset}, \eta_\mathrm{D}$ and $T_\mathrm{cycle}$ defined in the previous sections. To mitigate this source of noise, the qubit is thermalized by filtering the line on a broad frequency range until the IR domain (eccosorb filter) and by properly designing an electromagnetic shield composed of 3 interleaved screens in $\mu-$metal, copper and aluminium (see appendix).

%This contribution can be calculated since we know precisely the values of the different parameters : $p_\mathrm{reset} = 10^{-5}$, $T_1 = 37$ $\mathrm{\mu s}$, $p_\mathrm{eq} = 10^{-4}$, $T_\mathrm{cycle} = 12$ $\mathrm{\mu s}$, $\eta_\mathrm{D} = 0.84$. We obtain the dark count rate due to the qubit $\alpha_{\mathrm{qubit}} = 3$ $\mathrm{s^{-1}}$. This first contribution is made negligible  ($\approx 3 \%$ of the total dark count) thanks to a very good thermalization of the qubit and thanks to a good reset protocol made functional by using a state of the art quantum amplifier (TWPA) for the qubit readout. 

The second contribution is the spurious heating of the qubit by the pump tone $\alpha_\mathrm{4wm}$. This contribution is measured by applying a pump tone detuned from the four-wave mixing condition while measuring the equilibrium population of the qubit. As shown in Fig.\ref{fig2}b, in these conditions $p_\mathrm{eq} = 3\cdot 10^\mathrm{-4}$, a value included in the fluctuation interval of the equilibrium population. This contribution to the overall dark count rate is therefore considered negligible.    

%Based on the measurement of the qubit excited population, shown in Fig.\ref{fig2}b, when the pump tone is activated without buffer illumination, one can estimate that this contribution account for a rate $\alpha_\mathrm{4wm+} \approx 3$ $\mathrm{s^{-1}}$.  

%These arrangements make it possible to neglect the pump-induced heating, as shown in Fig.~\ref{fig3}c, where we measure the qubit state after a 10 $\mathrm{\mu s}$ pump pulse. When the pump frequency is detuned from the four-wave mixing state, the qubit is in its equilibrium population $p_\mathrm{eq}$. 

\begin{figure}[h]
  \includegraphics[width=\columnwidth]{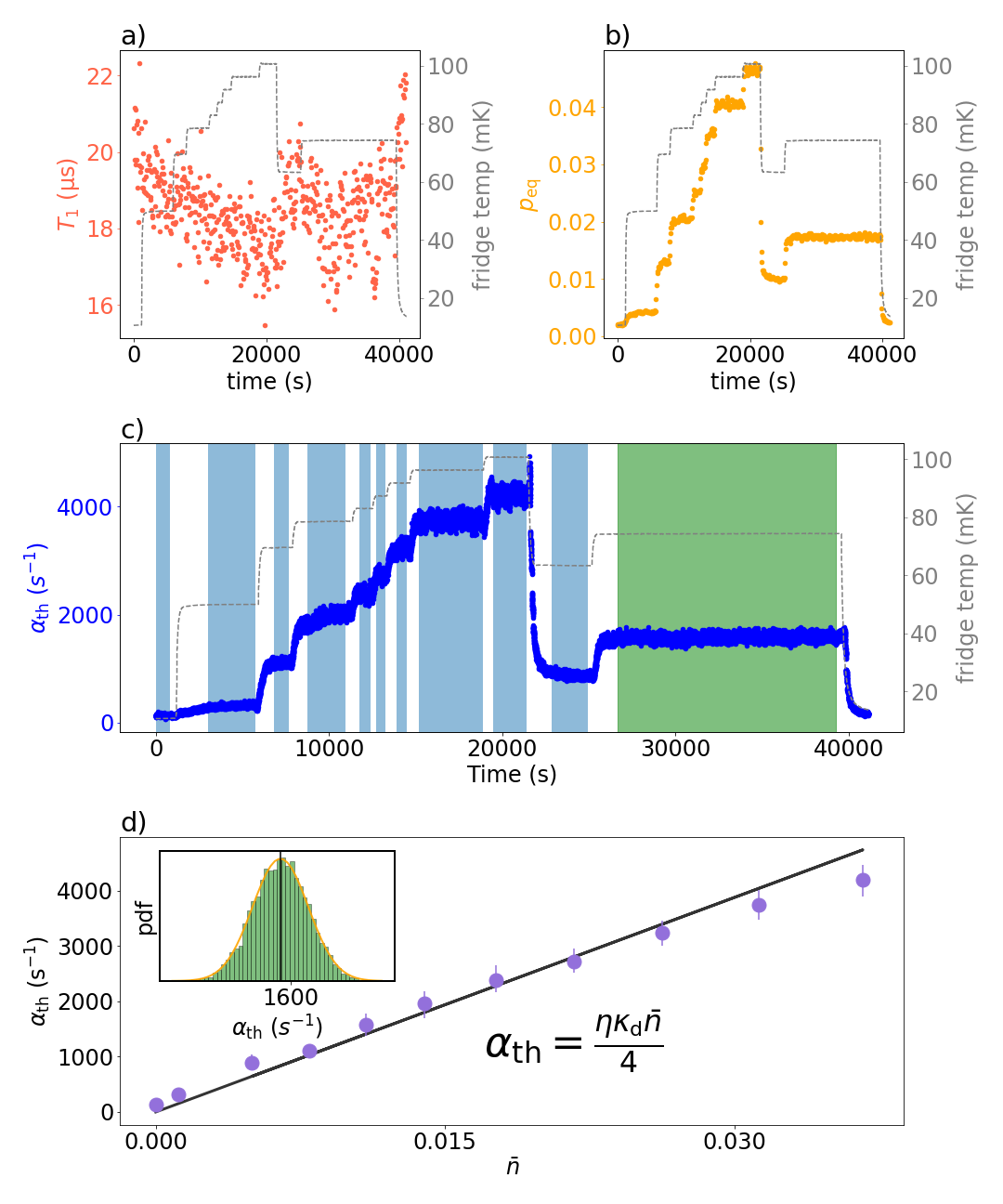}
  \caption{\label{fig4}
  Johnson-Nyquist law demonstrated with the SMPD.
  Qubit relaxation time $T_\mathrm{1}$ a) and qubit equilibrium population $p_\mathrm{eq}$ b) taken every minute. These values are used to determine the value of $\alpha_\mathrm{qubit}$. The fridge temperature is shown in dashed grey lines.
  c) Thermal dark count rate $\alpha_\mathrm{th} = \alpha - \alpha_\mathrm{qubit}$ for different refrigerator temperature shown in dashed grey lines. Each point corresponds to the average dark count rate over $10^{5}$ cycles, the light blue areas correspond to the data selected to extract the averages by avoiding the transient regime. 
  d) Relation between the mode population $\Bar{n}_\mathrm{b}$, calculated from the refrigerator temperature, and the thermal dark count rate $\alpha_\mathrm{th}$. Each point (purple) corresponds to the average of the colored areas of (c), an example of distribution is given in inset for the green zone. The solid line correspond to the Johnson Nyquist relation where $\eta$ is the SMPD efficiency and $\kappa_\mathrm{d}$ the SMPD bandwith.  
  }
\end{figure}

The third contribution $\alpha_\mathrm{th}$ is due to the presence of spurious photons in the input transmission lines. The integration of the mean number of photons per mode in the buffer input line $\Bar{n}_b$ over the linewidth of the detector $\kappa_\mathrm{d}$ gives the corresponding dark count rate $\alpha_\mathrm{th} = 80$ $\mathrm{s^{-1}}$. At the cryostat base temperature (10mK), this contribution should be negligible as the Planck law would predict an average number of photons per mode $\bar{n}_\mathrm{b} = 3 \cdot 10^\mathrm{-15}$; however, it is notoriously difficult to thermalize the microwave field at such low temperatures. Based on a Johnson-Nyquist description of thermal noise, we can derive the explicit relation \cite{balembois_magnetic_2023} (see Appendix F):

%To prove this hypothesis, we compare the dynamics of $\alpha_\mathrm{th}$ as a function of refrigerator temperature with the theoretical Johnson-Nyquist noise given by the following formula.

%This thermal noise is well understood in the framework of the Johnson-Nyquist formula in electronics circuits. We extrapolated this law in our experiment to find a quantitative relationship between the thermal photons and the dark count rate.

\begin{equation}
    \alpha_\mathrm{th}=  \frac{\kappa_\mathrm{d}}{4} \eta \Bar{n}_\mathrm{b},
\end{equation}

To verify the validity of this relation, we measure the thermal dark count as the function of a well defined mode population i.e when the refrigerator temperature is superior to 40 mK. This temperature is measured by a thermometer anchored to the mixing chamber plate. We acquired the dark count rate $\alpha$ in a range from 10 mK to 100 mK, waiting for the end of the transient regime each time. Since the heating is not selective, all parts of the chip are affected including the transmon qubit which causes an increase in its equilibrium population $p_\mathrm{eq}$ and a decrease in its relaxation time $T_\mathrm{1}$. To take these effects into account, these two quantities are measured every minutes during the experiment (Fig.~\ref{fig4}a,b), giving minute-by-minute monitoring of $\alpha_\mathrm{qubit}$. Note that the relaxation time is now of the order of 20 $\mathrm{\mu s}$ compared with 37 $\mathrm{\mu s}$ in the previous sections. This decrease appeared for an unknown reason after a few cool-down cycles. 

The thermal dark count rate $\alpha_\mathrm{th} = \alpha - \alpha_\mathrm{qubit}$ is plotted as a function of time (see Fig.~\ref{fig4}c) and with the respect to the thermal photon population $\Bar{n}_\mathrm{b}$ calculated from the refrigerator temperature (see Fig.~\ref{fig4}d). The relationship between $\alpha_\mathrm{th}$ and $\Bar{n}_\mathrm{b}$ is linear with a slope of $\eta\kappa_\mathrm{d}/4 $, therefore validating Eq. (5). 

As explained earlier, at 10 mK, $\Bar{n}_\mathrm{b}$ is decoupled from the refrigerator temperature, but we can estimate an equivalent electromagnetic temperature from the dark count measurement. At 10 mK, the measured dark count is $\alpha = 85$ $\mathrm{s^{-1}}$ (see Fig.~\ref{fig3}c), as we evaluate $\alpha_\mathrm{qubit}$ to 5 $\mathrm{s^{-1}}$, the equivalent $\alpha_\mathrm{th}$ is $80$ $\mathrm{s^{-1}}$, corresponding to $\Bar{n}_\mathrm{b} = 6.5 \cdot 10^\mathrm{-5}$. By using Eq. (5), the equivalent electromagnetic temperature of the input line is 35 mK.

In principe, one could further improve the SMPD performances by improving the attenuation of the lines. However, the temperature of the microwave radiations is challenging to reduce arbitrarily close to the cryostat base temperature as it requires a large amount of attenuator that are well thermally anchored. 

\section{Conclusion}

In conclusion, we have demonstrated the operation of a single microwave photon detector based on a four-wave mixing process.  

The efficiency of the device reaches 0.43 and is quantitatively understood from the contributions of the detector duty cycle, the qubit ability to store an excitation, the qubit readout, and the 4 waves mixing efficiency. It can be improved in future devices with longer transmon relaxation times (see Appendix H), noting that qubit $T_\mathrm{1}$s up to several hundred microseconds have been demonstrated \cite{place_new_2021,wang_towards_2022} 

The second key quantity studied in this article is the dark count rate. We have demonstrated that most of these false positives events are caused by spurious photons due to the electromagnetic temperature of the line. A direct way of lowering the dark count would be to reduce the detector bandwidth so that it matches the bandwidth of the measured system. 

Utilizing these metrics, the power sensitivity of the SMPD is determined as $\mathcal{S} = \hbar\omega\frac{\sqrt{\alpha}}{\eta} =  10^{-22}$ $\mathrm{W/\sqrt{Hz}}$. We have also verified the direct relation between the count rates and the thermal occupation of the lines, opening the way to using the SMPD as an absolute thermometer in the 10-100mK range. 

Even though further improvements of the device performance are desirable in the future, its high sensitivity already enabled new experiments, such as single-spin Electron Spin Resonance (ESR) spectroscopy \cite{wang_single_2023}, as well as proof-of-principle axion search.

\subsection*{Acknowledgements}
{We acknowledge technical support from P.~S\'enat, D. Duet, P.-F.~Orfila and S.~Delprat, and are grateful for fruitful discussions within the Quantronics group. We acknowledge support from the Agence Nationale de la Recherche (ANR) through the DARKWADOR (ANR-19-CE47-0004) projects. We acknowledge support of the R\'egion Ile-de-France through the DIM SIRTEQ (REIMIC project), from the AIDAS virtual joint laboratory, and from the France 2030 plan under the ANR-22-PETQ-0003 grant. This project has received funding from the European Research Council under grant no. 101042315 (INGENIOUS). We acknowledge IARPA and Lincoln Labs for providing the Josephson Traveling-Wave Parametric Amplifier. }

\clearpage
\onecolumngrid
\section*{Appendices}
\twocolumngrid
\appendix

%\title{Supplementary informations: Cyclically operated Single Microwave Photon Counter with $10^\mathrm{-22}$ $\mathrm{W/\sqrt{Hz}}$ sensitivity.}

\section{Experimental parameters of the device}

\begin{table}[!tbh]
\begin{tabular}{|c c c c|}
\hline
\textbf{Qubit} &&& \\
\hline
$\omega_\mathrm{q}/2\pi$ &&& $6.184$ GHz \\
$E_\mathrm{c}/4\pi$ &&& $\sim 240$ MHz \\
$\chi_\mathrm{b}/2\pi$ &&& $5.2$ MHz \\
$\chi_\mathrm{w}/2\pi$ &&& $18.8$ MHz \\
$T_\mathrm{1}$			&&& $\sim 37 \,\mu$s \\
$T_\mathrm{2}$			&&& $\sim 56 \,\mu$s \\
$p_\mathrm{eq}$   &&& $\sim 2\cdot 10^{-4}$ \\

\hline
\textbf{Buffer}  & top of arch & unbiased & at resonance  \\
\hline
$\omega_\mathrm{b}/2\pi$ 		&$7.005$ GHz& $6.979$ GHz &$6.824$ GHz   \\
$\kappa_\mathrm{b,ext}/2\pi$ 	&$0.152$ MHz& $0.172$ MHz &$2.95$ MHz	  \\
$\kappa_\mathrm{b,int}/2\pi$ 	& $0.100$ MHz& $0.028$ MHz & nc  \\
\hline
\textbf{Waste} &&&\\
\hline
$\omega_\mathrm{w}/2\pi$ &&& $7.704$ GHz \\
$\kappa_\mathrm{w,ext}/2\pi$ &&& $1.72$ MHz \\
$\kappa_\mathrm{w,int}/2\pi$ &&& $0.11$ MHz \\

\hline
\textbf{Filters}  & & &  \\
\hline
$\omega_\mathrm{Pb}/2\pi$ &&& $6.824$ GHz \\
$\kappa_\mathrm{Pb}/2\pi$ &&& $84.2$ MHz \\
$\omega_\mathrm{Pw}/2\pi$ &&& $7.620$ GHz \\
$\kappa_\mathrm{Pw}/2\pi$ &&& $180$ MHz \\
\hline

\end{tabular}
\end{table}

The table summarises the various parameters of the device. The parameter $\omega_\mathrm{Pw}$ ($\omega_\mathrm{Pb}$) is the frequency of the waste (buffer) Purcell filter. $\kappa_\mathrm{Pb}, \kappa_\mathrm{Pw}$ are the filters bandwidth. $E_\mathrm{c}/2$ is the anharmonicity of the qubit. $\kappa_\mathrm{b,ext} (\kappa_\mathrm{b,int})$ is the coupling (internal) losses of the buffer resonator and $\kappa_\mathrm{w,ext} (\kappa_\mathrm{w,int})$ is the coupling (internal) losses of the waste resonator. 

\section{Other operating points}

In the main text, we present the SMPD for the specific buffer frequency $\omega_\mathrm{b}/2\pi = 6.979$ GHz. In this section, we present the dark count and the efficiency of the detector at two other buffer frequencies: $\omega_\mathrm{b}/2\pi = 6.9697$ GHz and $\omega_\mathrm{b}/2\pi = 6.9549$ GHz. 

\begin{figure}[t!]
  \includegraphics[width=\columnwidth]{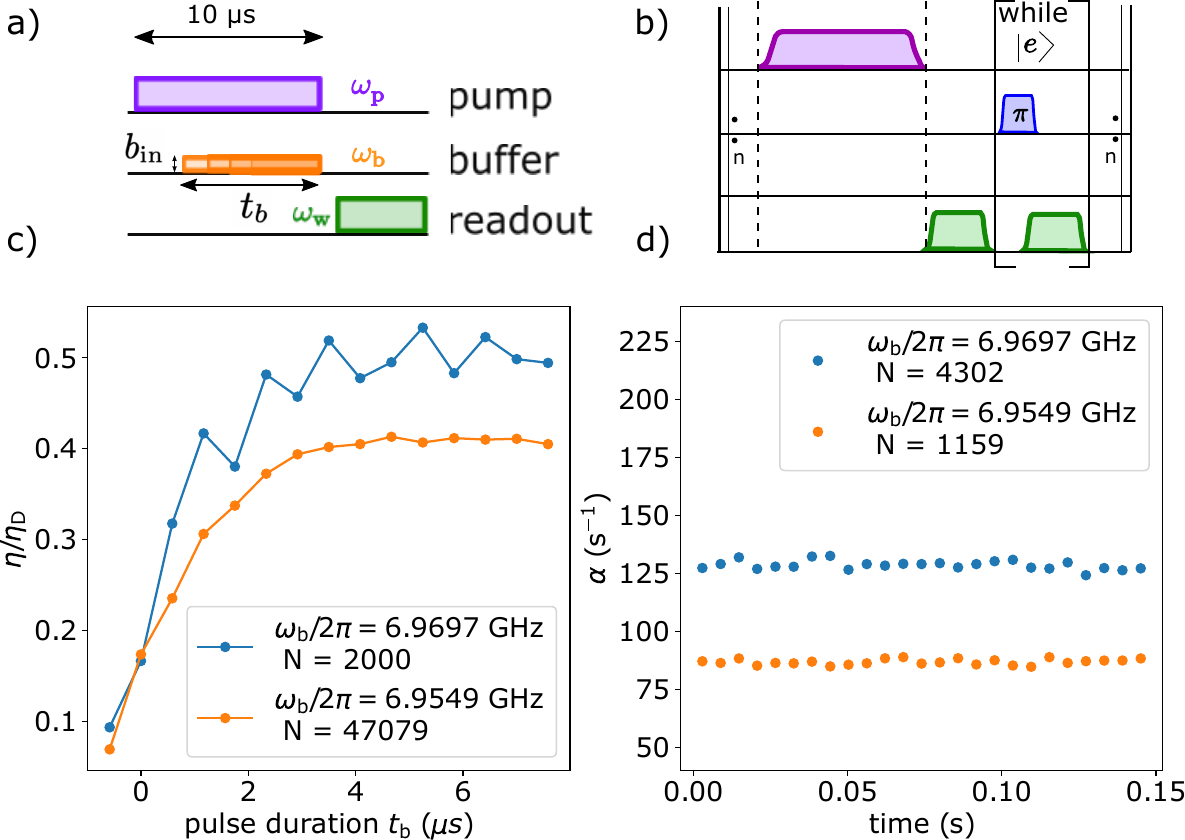}
  \caption{\label{fig:extended_fig1} a) Pulse sequence used to measure efficiency. A pump tone (purple pulse) of duration 10 $\mathrm{\mu s}$ is sent to the qubit mode, and an incoming coherent wave packet (orange pulse) of duration $t_\mathrm{b}$ and amplitude $b_\mathrm{in}$ is applied to the buffer. The qubit excited state population $p_\mathrm{e}$ is detected by the waste (green pulse). b) Cycle similar to the main text repeated n = 12500 times per experiment. c) Efficiency $\eta/\eta_\mathrm{D}$ as a function of wave packet duration $t_\mathrm{b}$, measured with sequence a). Each point is averaged N times. d) Dark count measured over n = 12500 cycles of $t_\mathrm{cycle} = 11.8$ $\mathrm{\mu s}$. Each point corresponds to the average of k = 500 cycles. The experiment is repeated N times and averaged.      
  }
\end{figure}

The efficiency is measured by sending a coherent wave packet of amplitude $b_\mathrm{in}$ to the buffer for a time $t_\mathrm{b}$, while the four-wave mixing is activated by the pump tone (see Fig. \ref{fig:extended_fig1}a). The photon flux corresponding to the amplitude $b_\mathrm{in}$ is calibrated using a Ramsey experiment perturbed by a field sent to the buffer (see \cite{balembois_magnetic_2023} for more details). For each time $t_\mathrm{b}$ we measure the ratio between the number of photons sent and the qubit excited population, which gives the detector efficiency without the duty cycle contribution: $\eta/\eta_\mathrm{D}$. As shown in Fig. \ref{fig:extended_fig1}c, due to the buffer bandwidth, the efficiency grows with $t_\mathrm{b}$ until it reaches an asymptote corresponding to $\eta/\eta_\mathrm{D}$. Taking into account the duty cycle, $\eta_\mathrm{D} = 0.84$, we get $\eta = 0.41$ for $\omega_\mathrm{b}/2\pi = 6.9697$ GHz and $\eta = 0.34$ for $\omega_\mathrm{b}/2\pi = 6.9549$ GHz. 

The dark count rate $\alpha$ is measured by repeating the detection cycle $N\cdot n$ times,  where N is the number of experiments, and $n=12500$ is the number of detection cycles of $t_\mathrm{cycle} = 11.8$ $\mathrm{\mu s}$ per experiment. For each experiment, the number of clicks per second is averaged over $k=500$ cycles and plotted as a function of time in Fig. \ref{fig:extended_fig1}d. We get $\alpha = 125$ $\mathrm{s^{-1}}$ for $\omega_\mathrm{b}/2\pi = 6.9697$ GHz and $\eta = 90$ $\mathrm{s^{-1}}$ for $\omega_\mathrm{b}/2\pi = 6.9549$ GHz. 

These figures give a sensitivity of $\mathcal{S} \approx 1.2\cdot 10^{-22}$ $\mathrm{W/\sqrt{Hz}}$ for both buffer frequencies, comparable to the sensitivity shown in the main text. 

\section{Noise Equivalent Power (NEP)}

The NEP is defined as the minimum detectable power with an signal-to-noise ratio (SNR) of 1 for a certain integration time $t$. This quantity, expressed in $\mathrm{W/\sqrt{Hz}}$, provides a good representation of the absolute sensitivity of the SMDP.

We will first write the signal-to-noise ratio considering that the detected signal is provided by a continuous tone of power $P$, at resonance with the buffer resonator and with a Poissonian noise. When this microwave tone is turning ON, the number of photon impinges the detector for a time $t$ is $Pt/\hbar\omega_\mathrm{b}$. Due to the dark count rate and the efficiency, the number of clicks given by the detector is $S_\mathrm{ON} = \eta Pt/\hbar\omega_\mathrm{b} + \alpha t$. On the contrary, when the microwave tone is OFF, the signal integrated by the detector for a time $t$ is $S_\mathrm{OFF} = \alpha t$. 

The signal of interest is $S_\mathrm{int} = S_\mathrm{ON}-S_\mathrm{OFF} = \eta Pt/\hbar\omega_\mathrm{b}$. As all the distributions are Poissonian, the associated noise is $N_\mathrm{int} = \sqrt{S_\mathrm{ON}+S_\mathrm{OFF}}$. Assuming that the dark count is perfectly known we can reduce the expression of the noise to $N_\mathrm{int} = \sqrt{S_\mathrm{ON}}$.

The SNR of the detection is then given by: 

\begin{equation}
    \mathrm{SNR} = \frac{\eta Pt/\hbar\omega_\mathrm{b}}{\sqrt{Pt/\hbar\omega_\mathrm{b} + \alpha t}}.
\end{equation}
 
The NEP is given by the power  $P$ corresponding to  $\mathrm{SNR}=1$ yielding to:

\begin{equation}
    \mathrm{NEP} = \frac{\hbar\omega_\mathrm{b}(1+\sqrt{1+4t\alpha})}{2t\eta}.
\end{equation}

In this paper, the integration time is set such as $\sqrt{\alpha t}\gg1$ reducing the NEP to: 

\begin{equation}
   \mathrm{NEP} = \hbar\omega_\mathrm{b}\frac{\sqrt{\alpha}}{\eta\sqrt{t}}.
    \label{eq:SMDP_Nep}
\end{equation}

\section{Two coupled cavities model}

The detector response can be modeled by considering that the buffer and waste resonator are coupled with a constant $\mathcal{G}=-\xi_p\sqrt{\chi_{qb}\chi_{qw}}$ due to the 4-wave parametric process involving the qubit, where $\xi_p$ is the pump amplitude in units of square root of photons and $\chi_{qb} (\chi_{qw})$ the dispersive coupling of the buffer (waste) resonator to the qubit.
We can write down the system of coupled equations for the buffer and waste intra-resonator fields $\nu$ and $\beta$, assuming that the resonators are lossless:

\begin{equation}\label{eq:input_outpout_beta}
    \dot{\nu}= -i\delta_b \nu -i \mathcal{G}\beta-\frac{\kappa_b}{2}+\sqrt{\kappa_b} \nu_{in}
\end{equation}

\begin{equation}
    \dot{\beta}= -i\delta_w \beta -i \mathcal{G}^*\nu-\frac{\kappa_w}{2}+\sqrt{\kappa_w} \beta_{in}
\end{equation}

where $\delta_b$ and $\delta_w$ are the buffer and waste frequencies in the frame rotating  at the probing frequency and $\nu_{in}$, $\beta_{in}$ are the respective input field amplitudes. Now using the relation between the intra-resonator fields and the input and output flux $\sqrt{\kappa_b}\nu=\nu_{in}+\nu_{out}$, from the equilibrium solution of the coupled system we can extract the transmission coefficient $\left|S_{21}\right|^2=\left|\beta_{out}/\nu_{in}\right|^2$. Assuming zero input flux on the waste this leads to:
\begin{equation}
    \left|\mathrm{S_{21}}\right|^2=\left|\frac{2\mathrm{\xi_p}\sqrt{\mathrm{\kappa_b\kappa_w\chi_b\chi_w}}}{-4\delta_\mathrm{b}\delta_\mathrm{w} 
    + 2i\delta_\mathrm{b}\mathrm{\kappa_w} 
    +2i\delta_\mathrm{w}\mathrm{\kappa_b} 
    +\mathrm{\kappa_b\kappa_w} 
    +\mathrm{\chi_b\chi_w \xi_p^2}}\right|^2,
\end{equation}

or with the respect to the cooperativity $C = 4\frac{\chi_\mathrm{b}\chi_\mathrm{w}}{\kappa_\mathrm{w}\kappa_\mathrm{b}}$: 

\begin{equation}\label{eq:transmission_coeff}
    \left|\mathrm{S_{21}}\right|^2=\frac{4C}{\left|-\frac{4\delta_\mathrm{b}\delta_\mathrm{w}}{\kappa_\mathrm{b}\kappa_\mathrm{w}} 
    + 2i\frac{\delta_\mathrm{b}}{\kappa_\mathrm{b}}
    +2i\frac{\delta_\mathrm{w}}{\kappa_\mathrm{w}}
    + 1  
    +C \right|^2}.
\end{equation}

This expression can be directly related to the detector efficiency $\eta_\mathrm{4wm}$ as the input photon frequency is varied. The full-width-half-maximum of $\left|\mathrm{S_{21}}\right|^2$:

\begin{equation}\label{eq:SMPD_BW}
\kappa_\mathrm{d}=\sqrt{2}\sqrt{\sqrt{\kappa_\mathrm{b}^2\kappa_\mathrm{w}^2+\left(\frac{\kappa_\mathrm{b}-\kappa_\mathrm{w}}{2}\right)^4}-\left(\frac{\kappa_\mathrm{b}-\kappa_\mathrm{w}}{2}\right)^2},
\end{equation}

gives the bandwidth of the detector. 

\section{Transmission coefficient in presence of buffer resonator losses}

The effect of the buffer resonator internal losses on the efficiency $\eta_\mathrm{4wm}$ can be evaluated by inserting $\kappa_\mathrm{tot} = \kappa_\mathrm{b,int} + \kappa_\mathrm{b,ext}$ (internal and coupling losses) in Eq \eqref{eq:input_outpout_beta}. Assuming that the incoming photon and the pump frequencies are optimally tuned (i.e. $\delta_\mathrm{b} = \delta_\mathrm{w} = 0$), the transmission coefficient becomes: 

\begin{equation}
    \left|\mathrm{S_{21}}\right|^2 = \frac{4\mathcal{C}}{(\frac{\kappa_\mathrm{b,int}}{\kappa_\mathrm{b,ext}}+1+\mathcal{C})^2}.
\end{equation}

The maximization of $\left|\mathrm{S_{21}}\right|^2$ with the respect of the cooperativity $C$ yields to: 

\begin{equation}
    \left|\mathrm{S_{21}}\right|^2 \le \frac{1}{1+\frac{\kappa_\mathrm{b,int}}{\kappa_\mathrm{b,ext}}}.
\end{equation}

The internal losses of the buffer resonator $\kappa_\mathrm{b,int}/2\pi = 0.028$ MHz combined with the external (coupling) losses $\kappa_\mathrm{b,ext}/2\pi = 0.172$ MHz gives the maximum efficiency of the main text $\eta_\mathrm{4wm} = 0.86$.

\section{Johnson-Nyquist noise}

The thermal noise detected in the main text is described by a Johnson-Nyquist noise. In the classical framework, the noise power  is expressed as a function of the detector bandwidth $\Delta f$ and the temperature $T$ of the experiment as: $P_\mathrm{th} = k_\mathrm{b}T\Delta f$. In the quantum regime relevant for our  experiment performed at low temperature 10 mK ($k_\mathrm{b}T \ll \hbar\omega_\mathrm{b}$),   the average energy provided by the modes is given by  Bose-Einsten statistics such as: $ k_\mathrm{b}T \rightarrow  \hbar\omega\Bar{n}_\mathrm{b} $ with $\Bar{n}_\mathrm{b} = 1/(e^{\hbar\omega_\mathrm/k_\mathrm{b}T}-1)$ the number of photons per mode. The expression describing the flux of thermal photons per second is then: 

\begin{equation}
    \frac{P_\mathrm{th}}{\hbar\omega_\mathrm{b}} = \Bar{n}_\mathrm{b}\Delta f
    \label{flux_thermal_photon}
\end{equation}

To extract the extra number of clicks $\alpha_\mathrm{th}$ induced by this photon flux, we must take into account its conversion efficiency, which depends on the total detector efficiency $\eta$, but also on its frequency detuning with the buffer resonator. In the limit where $\kappa_\mathrm{b}\ll\kappa_\mathrm{w}$, we can consider that the conversion efficiency  $|S_\mathrm{21}|^2(f)$ is  given by  a Lorentzian  function   centered around $f_\mathrm{b} = \omega_\mathrm{b}/2\pi$ with a FWHM $\kappa_\mathrm{d}/2\pi$ . This assumption yields  the total number of extra clicks during  a detection window: 

\begin{equation}
    \alpha_\mathrm{th} = \int_{-\infty}^{+\infty}\frac{\bar{n}_\mathrm{b}\eta}{1+(\frac{f-f_\mathrm{b}}{\kappa_\mathrm{d}/(4\pi)})^2}df
\end{equation}

\begin{equation}
    \alpha_\mathrm{th} = \frac{\bar{n}_\mathrm{b}\eta\kappa_\mathrm{d}}{4}
\end{equation}

\section{Otimization of the detection time $T_\mathrm{D}$}

In the main text, we define the qubit efficiency as $\eta_\mathrm{qubit} = \frac{T_\mathrm{1}}{T_\mathrm{d}}(1-e^{-T_\mathrm{d} / T_\mathrm{1}})$. This expression can approach 1 closely, by shortening the detection window, thereby diminishing the ratio $T_\mathrm{d}/T_\mathrm{1}$. Nevertheless, reducing $T_\mathrm{d}$ also decreases the duty cycle $\eta_\mathrm{D}$. One has thus to find a trade-off between $\eta_\mathrm{D}$ and $\eta_\mathrm{qubit}$ by maximizing the product:

\begin{equation}
    \eta_\mathrm{D}\eta_\mathrm{qubit} = \frac{T_\mathrm{1}}{T_\mathrm{m}+T_\mathrm{r}+T_\mathrm{d}}(1-e^{-T_\mathrm{d}/T_\mathrm{1}}).
\end{equation}

In the   limit where   $T_\mathrm{m}+T_\mathrm{r}\ll T_\mathrm{d}\ll T_\mathrm{1}$ , the product $\eta_\mathrm{D}\eta_\mathrm{qubit} $ takes the simple form: 

\begin{equation}
    \eta_\mathrm{D}\eta_\mathrm{qubit} \approx \left(1-\frac{T_\mathrm{m} + T_\mathrm{r}}{T_\mathrm{d}}\right)\left(1-\frac{T_\mathrm{d}}{2T_\mathrm{1}}\right).
\end{equation}

The optimal detection window  is then equal to :

\begin{equation}\label{eq:detection time}
    T_\mathrm{d} \approx \sqrt{2(T_\mathrm{m}+T_\mathrm{r})T_\mathrm{1}}.
\end{equation}

Taking into account the parameters of our system, we choose $T_\mathrm{d} = 10$ $\mathrm{\mu s}$. 

\section{Effect of qubit relaxation time improvement on the SMPD sensitivity}

Within the main text, we assert that augmenting the qubit's $T_\mathrm{1}$ stands as a pivotal factor in advancing the detector's sensitivity to a greater extent.  Indeed, an extended $T_\mathrm{1}$ inherently enhances the efficiency of the qubit, $\eta_\mathrm{qubit}$. Furthermore, under the assumption of a constant $T_\mathrm{r}+ T_\mathrm{m}$, from eq. \eqref{eq:detection time}, it becomes apparent that the optimal detection time rises, thereby amplifying the duty cycle $\eta_\mathrm{D}$. An extended relaxation time contributes also to heightened readout efficiency, $\eta_\mathrm{RO}$. It becomes possible to increase the readout time, $T_\mathrm{r}$, without encountering adverse effects from relaxation events during measurement. This naturally enhances the distinction between the two states within the phase plane.

The qubit relaxation time also plays a role in the SMPD dark count rate, as it appears directly in $\alpha_{\mathrm{qubit}} = \frac{p_{\mathrm{eq}}}{T_\mathrm{1}}\eta_\mathrm{D} + \frac{p_{\mathrm{reset}}}{T_{\mathrm{cycle}}}$ and indirectly in $\alpha_\mathrm{th}= \frac{\kappa_\mathrm{d}}{4} \eta \Bar{n}_\mathrm{b}$ by the total efficiency $\eta$. 

Assuming that the $T_\mathrm{1}$ of the device presented in the main text is larger by an order of magnitude, it follows that $\eta_\mathrm{qubit}, \eta_\mathrm{RO}, \eta_\mathrm{D} \sim 1$ and $\eta \sim \eta_\mathrm{4wm}$. Concerning the dark count rate, $\alpha_\mathrm{qubit} \sim 0$ $\mathrm{s^{-1}}$ and $\alpha \sim \alpha_\mathrm{th}$. 

Under these conditions, we can estimate the new sensitivity would be $\mathcal{S}_\mathrm{high T_\mathrm{1}} = 6.8 \cdot 10^{-23}$ $\mathrm{W/\sqrt{Hz}}$ (to be compared with the actual sensitivity $\mathcal{S} = 10^{-22}$ $\mathrm{W/\sqrt{Hz}}$). 

To further improve the sensitivity of the detector, it's crucial to reduce the dark count rate due to the thermal population of the line. As mentioned in the main text, one option is to reduce the bandwidth of the detector to match the bandwidth of the system being measured. A detailed study of the dark count rate behaviour with the respect to the bandwith of the SMPD is given in \cite{balembois_magnetic_2023}.

\section{Twin SMPD device}

In the main text we present the characterisation of the first working device (SMPD1) with a sensitivity $\mathcal{S} = 10^\mathrm{-22}$ $\mathrm{W/\sqrt{Hz}}$. A twin device (SMPD2) was fabricated a few months later to be used in a spin detection experiment \cite{wang_single_2023}. Its resonator frequencies ($\omega_\mathrm{b}/2\pi = 7.459 $ $\mathrm{GHz}$ unbiased, $\omega_\mathrm{w}/2\pi = 8. 004 $ $\mathrm{GHz}$) and qubit frequency ($\omega_\mathrm{q}/2\pi = 6.193 $ $\mathrm{GHz}$) were chosen to optimise the detector performances around 7.3 $\mathrm{GHz}$. The qubit has a smaller lifetime, $T_\mathrm{1} = 15$ $\mathrm{\mu s}$, and a comparable equilibrium population, $p_\mathrm{eq} = 2\cdot 10^\mathrm{-4}$. The measured efficiency $\eta = 0.32$ and the measured dark count rate $\alpha = 103$ $\mathrm{s^{-1}}$ give an absolute sensitivity $\mathcal{S} = 1.5 \cdot 10^{-22}$ $\mathrm{W/\sqrt{Hz}}$ at 7.3 GHz. The table below shows the efficiency and dark count budget of the SMPD2 and how it compares with the SMPD1.

\begin{table}[!tbh]
\begin{tabular}{|c c c c c|} 
\hline
Device & $\alpha_\mathrm{qubit} (\mathrm{s^{-1}})$ & $\alpha_\mathrm{4wm} (\mathrm{s^{-1}})$  & $\alpha_\mathrm{th} (\mathrm{s^{-1}})$ &\\
\hline
SMPD1 & 5 & / & 80 &\\
SMPD2 & 9 & 2 & 90 &\\
\hline
      & $\eta_\mathrm{D}$ & $\eta_\mathrm{RO}$ & $\eta_\mathrm{4wm}$ & $\eta_\mathrm{qubit}$\\
\hline

SMPD1 & 0.84 & 0.76 & 0.86 & 0.84\\
SMPD2 & 0.79 & 0.90 & 0.69 & 0.73\\
\hline
\end{tabular}
\end{table}

The decrease in $\eta_\mathrm{D}$ and $\eta_\mathrm{qubit}$ is due to the shorter $T_\mathrm{1}$, while the increase in $\eta_\mathrm{RO}$ is due to a cleaner calibration of the parametric amplifier. The SMPD2 buffer resonator present higher internal losses which translate into a smaller $\eta_\mathrm{4wm}$.

\section{Cryogenic setup}

\begin{figure*}[!tbh]
\includegraphics[width=0.85\textwidth]{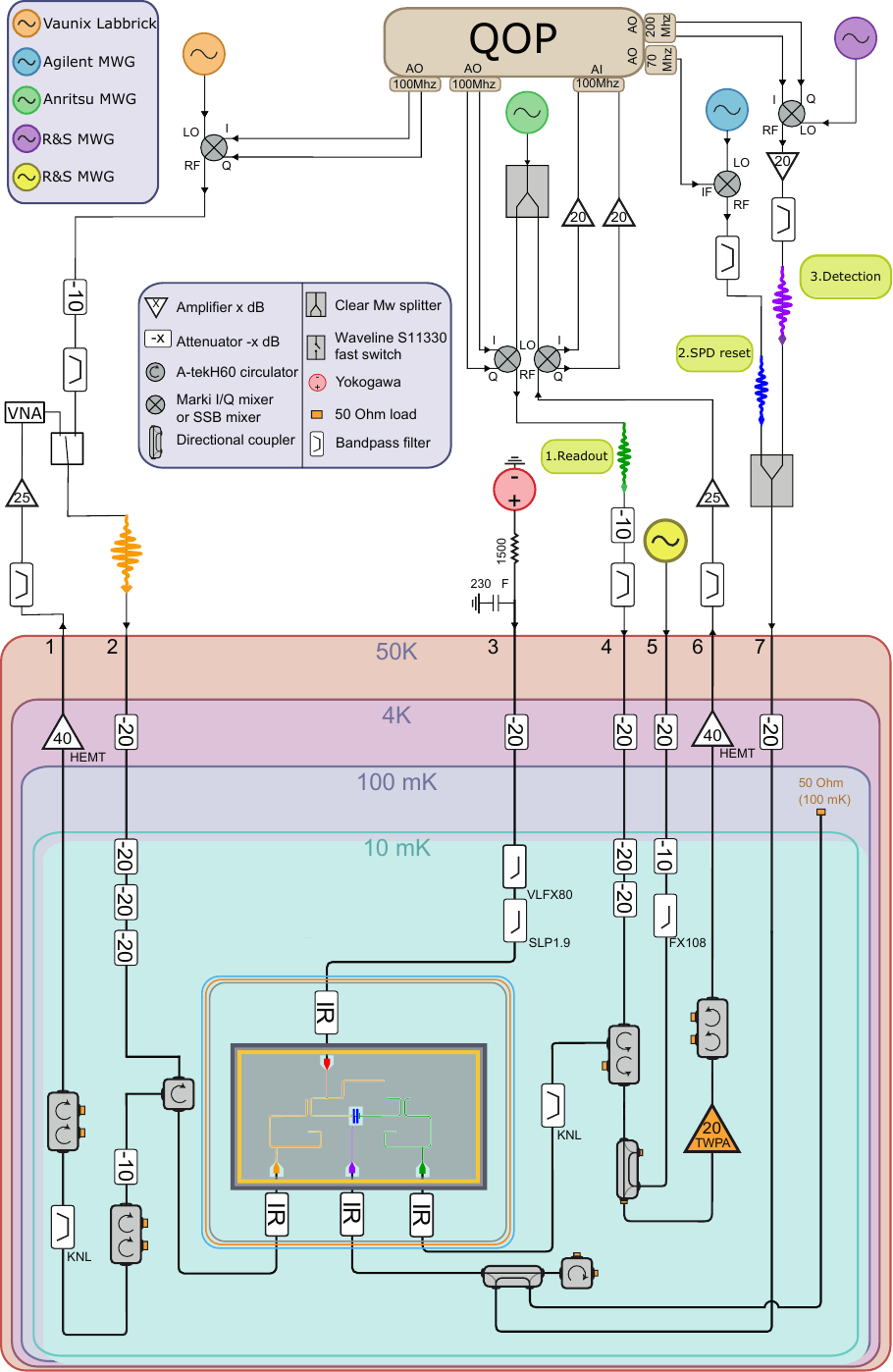}
\caption{
\textbf{Schematic of the setup.} Wiring and all the components used in this experiment at room temperature and cryogenic temperature are shown. 
}
\label{sub_fig1}
\end{figure*}

The figure \ref{sub_fig1} shows the wiring of the cryogenic setup used in this experiment. Lines 1 and 2 correspond to the detector input. They are used both to characterise the parameters of the buffer resonator ($\omega_\mathrm{b},\kappa_\mathrm{b}$) and to calibrate the detector efficiency by sending a well controlled number of photons. Line 3 is a DC flux bias line for tuning the SQUID inductance, which controls the frequency $\omega_\mathrm{b}$ of the buffer resonator. Lines 4, 5 and 6 are used to probe the waste resonator to perform a dispersive readout of the qubit. Line 7 corresponds to the qubit drive. 

\clearpage

\bibliography{NewGenSMPD}

\begin{thebibliography}{45}%
\makeatletter
\providecommand \@ifxundefined [1]{%
 \@ifx{#1\undefined}
}%
\providecommand \@ifnum [1]{%
 \ifnum #1\expandafter \@firstoftwo
 \else \expandafter \@secondoftwo
 \fi
}%
\providecommand \@ifx [1]{%
 \ifx #1\expandafter \@firstoftwo
 \else \expandafter \@secondoftwo
 \fi
}%
\providecommand \natexlab [1]{#1}%
\providecommand \enquote  [1]{``#1''}%
\providecommand \bibnamefont  [1]{#1}%
\providecommand \bibfnamefont [1]{#1}%
\providecommand \citenamefont [1]{#1}%
\providecommand \href@noop [0]{\@secondoftwo}%
\providecommand \href [0]{\begingroup \@sanitize@url \@href}%
\providecommand \@href[1]{\@@startlink{#1}\@@href}%
\providecommand \@@href[1]{\endgroup#1\@@endlink}%
\providecommand \@sanitize@url [0]{\catcode `\\12\catcode `\$12\catcode
  `\&12\catcode `\#12\catcode `\^12\catcode `\_12\catcode `\%12\relax}%
\providecommand \@@startlink[1]{}%
\providecommand \@@endlink[0]{}%
\providecommand \url  [0]{\begingroup\@sanitize@url \@url }%
\providecommand \@url [1]{\endgroup\@href {#1}{\urlprefix }}%
\providecommand \urlprefix  [0]{URL }%
\providecommand \Eprint [0]{\href }%
\providecommand \doibase [0]{http://dx.doi.org/}%
\providecommand \selectlanguage [0]{\@gobble}%
\providecommand \bibinfo  [0]{\@secondoftwo}%
\providecommand \bibfield  [0]{\@secondoftwo}%
\providecommand \translation [1]{[#1]}%
\providecommand \BibitemOpen [0]{}%
\providecommand \bibitemStop [0]{}%
\providecommand \bibitemNoStop [0]{.\EOS\space}%
\providecommand \EOS [0]{\spacefactor3000\relax}%
\providecommand \BibitemShut  [1]{\csname bibitem#1\endcsname}%
\let\auto@bib@innerbib\@empty
%</preamble>
\bibitem [{\citenamefont {Orrit}\ and\ \citenamefont
  {Bernard}(1990)}]{orrit_single_1990}%
  \BibitemOpen
  \bibfield  {author} {\bibinfo {author} {\bibfnamefont {M.}~\bibnamefont
  {Orrit}}\ and\ \bibinfo {author} {\bibfnamefont {J.}~\bibnamefont
  {Bernard}},\ }\href {\doibase 10.1103/PhysRevLett.65.2716} {\bibfield
  {journal} {\bibinfo  {journal} {Physical Review Letters}\ }\textbf {\bibinfo
  {volume} {65}},\ \bibinfo {pages} {2716} (\bibinfo {year}
  {1990})}\BibitemShut {NoStop}%
\bibitem [{\citenamefont {Klar}\ \emph {et~al.}(2000)\citenamefont {Klar},
  \citenamefont {Jakobs}, \citenamefont {Dyba}, \citenamefont {Egner},\ and\
  \citenamefont {Hell}}]{klar_fluorescence_2000}%
  \BibitemOpen
  \bibfield  {author} {\bibinfo {author} {\bibfnamefont {T.~A.}\ \bibnamefont
  {Klar}}, \bibinfo {author} {\bibfnamefont {S.}~\bibnamefont {Jakobs}},
  \bibinfo {author} {\bibfnamefont {M.}~\bibnamefont {Dyba}}, \bibinfo {author}
  {\bibfnamefont {A.}~\bibnamefont {Egner}}, \ and\ \bibinfo {author}
  {\bibfnamefont {S.~W.}\ \bibnamefont {Hell}},\ }\href {\doibase
  10.1073/pnas.97.15.8206} {\bibfield  {journal} {\bibinfo  {journal}
  {Proceedings of the National Academy of Sciences}\ }\textbf {\bibinfo
  {volume} {97}},\ \bibinfo {pages} {8206} (\bibinfo {year}
  {2000})}\BibitemShut {NoStop}%
\bibitem [{\citenamefont {Betzig}\ \emph {et~al.}(2006)\citenamefont {Betzig},
  \citenamefont {Patterson}, \citenamefont {Sougrat}, \citenamefont
  {Lindwasser}, \citenamefont {Olenych}, \citenamefont {Bonifacino},
  \citenamefont {Davidson}, \citenamefont {Lippincott-Schwartz},\ and\
  \citenamefont {Hess}}]{betzig_imaging_2006}%
  \BibitemOpen
  \bibfield  {author} {\bibinfo {author} {\bibfnamefont {E.}~\bibnamefont
  {Betzig}}, \bibinfo {author} {\bibfnamefont {G.~H.}\ \bibnamefont
  {Patterson}}, \bibinfo {author} {\bibfnamefont {R.}~\bibnamefont {Sougrat}},
  \bibinfo {author} {\bibfnamefont {O.~W.}\ \bibnamefont {Lindwasser}},
  \bibinfo {author} {\bibfnamefont {S.}~\bibnamefont {Olenych}}, \bibinfo
  {author} {\bibfnamefont {J.~S.}\ \bibnamefont {Bonifacino}}, \bibinfo
  {author} {\bibfnamefont {M.~W.}\ \bibnamefont {Davidson}}, \bibinfo {author}
  {\bibfnamefont {J.}~\bibnamefont {Lippincott-Schwartz}}, \ and\ \bibinfo
  {author} {\bibfnamefont {H.~F.}\ \bibnamefont {Hess}},\ }\href {\doibase
  10.1126/science.1127344} {\bibfield  {journal} {\bibinfo  {journal}
  {Science}\ }\textbf {\bibinfo {volume} {313}},\ \bibinfo {pages} {1642}
  (\bibinfo {year} {2006})}\BibitemShut {NoStop}%
\bibitem [{\citenamefont {Bruschini}\ \emph {et~al.}(2019)\citenamefont
  {Bruschini}, \citenamefont {Homulle}, \citenamefont {Antolovic},
  \citenamefont {Burri},\ and\ \citenamefont
  {Charbon}}]{bruschini_single-photon_2019}%
  \BibitemOpen
  \bibfield  {author} {\bibinfo {author} {\bibfnamefont {C.}~\bibnamefont
  {Bruschini}}, \bibinfo {author} {\bibfnamefont {H.}~\bibnamefont {Homulle}},
  \bibinfo {author} {\bibfnamefont {I.~M.}\ \bibnamefont {Antolovic}}, \bibinfo
  {author} {\bibfnamefont {S.}~\bibnamefont {Burri}}, \ and\ \bibinfo {author}
  {\bibfnamefont {E.}~\bibnamefont {Charbon}},\ }\href {\doibase
  10.1038/s41377-019-0191-5} {\bibfield  {journal} {\bibinfo  {journal} {Light:
  Science \& Applications}\ }\textbf {\bibinfo {volume} {8}},\ \bibinfo {pages}
  {87} (\bibinfo {year} {2019})}\BibitemShut {NoStop}%
\bibitem [{\citenamefont {Hadfield}()}]{hadfield_single-photon_2009}%
  \BibitemOpen
  \bibfield  {author} {\bibinfo {author} {\bibfnamefont {R.~H.}\ \bibnamefont
  {Hadfield}},\ }\href {\doibase 10.1038/nphoton.2009.230} {\ \textbf {\bibinfo
  {volume} {3}},\ \bibinfo {pages} {696}},\ \bibinfo {note} {number: 12
  Publisher: Nature Publishing Group}\BibitemShut {NoStop}%
\bibitem [{\citenamefont {Albertinale}\ \emph {et~al.}(2021)\citenamefont
  {Albertinale}, \citenamefont {Balembois}, \citenamefont {Billaud},
  \citenamefont {Ranjan}, \citenamefont {Flanigan}, \citenamefont {Schenkel},
  \citenamefont {Estève}, \citenamefont {Vion}, \citenamefont {Bertet},\ and\
  \citenamefont {Flurin}}]{albertinale_detecting_2021}%
  \BibitemOpen
  \bibfield  {author} {\bibinfo {author} {\bibfnamefont {E.}~\bibnamefont
  {Albertinale}}, \bibinfo {author} {\bibfnamefont {L.}~\bibnamefont
  {Balembois}}, \bibinfo {author} {\bibfnamefont {E.}~\bibnamefont {Billaud}},
  \bibinfo {author} {\bibfnamefont {V.}~\bibnamefont {Ranjan}}, \bibinfo
  {author} {\bibfnamefont {D.}~\bibnamefont {Flanigan}}, \bibinfo {author}
  {\bibfnamefont {T.}~\bibnamefont {Schenkel}}, \bibinfo {author}
  {\bibfnamefont {D.}~\bibnamefont {Estève}}, \bibinfo {author} {\bibfnamefont
  {D.}~\bibnamefont {Vion}}, \bibinfo {author} {\bibfnamefont {P.}~\bibnamefont
  {Bertet}}, \ and\ \bibinfo {author} {\bibfnamefont {E.}~\bibnamefont
  {Flurin}},\ }\href {\doibase 10.1038/s41586-021-04076-z} {\bibfield
  {journal} {\bibinfo  {journal} {Nature}\ }\textbf {\bibinfo {volume} {600}},\
  \bibinfo {pages} {434} (\bibinfo {year} {2021})}\BibitemShut {NoStop}%
\bibitem [{\citenamefont {Billaud}\ \emph {et~al.}()\citenamefont {Billaud},
  \citenamefont {Balembois}, \citenamefont {Dantec}, \citenamefont {Rančić},
  \citenamefont {Albertinale}, \citenamefont {Bertaina}, \citenamefont
  {Chanelière}, \citenamefont {Goldner}, \citenamefont {Estève},
  \citenamefont {Vion}, \citenamefont {Bertet},\ and\ \citenamefont
  {Flurin}}]{billaud_microwave_2022}%
  \BibitemOpen
  \bibfield  {author} {\bibinfo {author} {\bibfnamefont {E.}~\bibnamefont
  {Billaud}}, \bibinfo {author} {\bibfnamefont {L.}~\bibnamefont {Balembois}},
  \bibinfo {author} {\bibfnamefont {M.~L.}\ \bibnamefont {Dantec}}, \bibinfo
  {author} {\bibfnamefont {M.}~\bibnamefont {Rančić}}, \bibinfo {author}
  {\bibfnamefont {E.}~\bibnamefont {Albertinale}}, \bibinfo {author}
  {\bibfnamefont {S.}~\bibnamefont {Bertaina}}, \bibinfo {author}
  {\bibfnamefont {T.}~\bibnamefont {Chanelière}}, \bibinfo {author}
  {\bibfnamefont {P.}~\bibnamefont {Goldner}}, \bibinfo {author} {\bibfnamefont
  {D.}~\bibnamefont {Estève}}, \bibinfo {author} {\bibfnamefont
  {D.}~\bibnamefont {Vion}}, \bibinfo {author} {\bibfnamefont {P.}~\bibnamefont
  {Bertet}}, \ and\ \bibinfo {author} {\bibfnamefont {E.}~\bibnamefont
  {Flurin}},\ }\href {http://arxiv.org/abs/2208.13586} {\enquote {\bibinfo
  {title} {Microwave fluorescence detection of spin echoes},}\ }\Eprint
  {http://arxiv.org/abs/2208.13586 [quant-ph]} {2208.13586 [quant-ph]}
  \BibitemShut {NoStop}%
\bibitem [{\citenamefont {Wang}\ \emph {et~al.}(2023)\citenamefont {Wang},
  \citenamefont {Balembois}, \citenamefont {Rančić}, \citenamefont {Billaud},
  \citenamefont {Dantec}, \citenamefont {Ferrier}, \citenamefont {Goldner},
  \citenamefont {Bertaina}, \citenamefont {Chanelière}, \citenamefont
  {Estève}, \citenamefont {Vion}, \citenamefont {Bertet},\ and\ \citenamefont
  {Flurin}}]{wang_single_2023}%
  \BibitemOpen
  \bibfield  {author} {\bibinfo {author} {\bibfnamefont {Z.}~\bibnamefont
  {Wang}}, \bibinfo {author} {\bibfnamefont {L.}~\bibnamefont {Balembois}},
  \bibinfo {author} {\bibfnamefont {M.}~\bibnamefont {Rančić}}, \bibinfo
  {author} {\bibfnamefont {E.}~\bibnamefont {Billaud}}, \bibinfo {author}
  {\bibfnamefont {M.~L.}\ \bibnamefont {Dantec}}, \bibinfo {author}
  {\bibfnamefont {A.}~\bibnamefont {Ferrier}}, \bibinfo {author} {\bibfnamefont
  {P.}~\bibnamefont {Goldner}}, \bibinfo {author} {\bibfnamefont
  {S.}~\bibnamefont {Bertaina}}, \bibinfo {author} {\bibfnamefont
  {T.}~\bibnamefont {Chanelière}}, \bibinfo {author} {\bibfnamefont
  {D.}~\bibnamefont {Estève}}, \bibinfo {author} {\bibfnamefont
  {D.}~\bibnamefont {Vion}}, \bibinfo {author} {\bibfnamefont {P.}~\bibnamefont
  {Bertet}}, \ and\ \bibinfo {author} {\bibfnamefont {E.}~\bibnamefont
  {Flurin}},\ }\href {\doibase https://doi.org/10.1038/s41586-023-06097-2}
  {\bibfield  {journal} {\bibinfo  {journal} {Nature 619, 276–281 (2023)}\ }
  (\bibinfo {year} {2023}),\
  https://doi.org/10.1038/s41586-023-06097-2}\BibitemShut {NoStop}%
\bibitem [{\citenamefont {Lamoreaux}\ \emph {et~al.}(2013)\citenamefont
  {Lamoreaux}, \citenamefont {van Bibber}, \citenamefont {Lehnert},\ and\
  \citenamefont {Carosi}}]{lamoreaux_analysis_2013}%
  \BibitemOpen
  \bibfield  {author} {\bibinfo {author} {\bibfnamefont {S.~K.}\ \bibnamefont
  {Lamoreaux}}, \bibinfo {author} {\bibfnamefont {K.~A.}\ \bibnamefont {van
  Bibber}}, \bibinfo {author} {\bibfnamefont {K.~W.}\ \bibnamefont {Lehnert}},
  \ and\ \bibinfo {author} {\bibfnamefont {G.}~\bibnamefont {Carosi}},\ }\href
  {\doibase 10.1103/PhysRevD.88.035020} {\bibfield  {journal} {\bibinfo
  {journal} {Physical Review D}\ }\textbf {\bibinfo {volume} {88}},\ \bibinfo
  {pages} {035020} (\bibinfo {year} {2013})}\BibitemShut {NoStop}%
\bibitem [{\citenamefont {Dixit}\ \emph {et~al.}()\citenamefont {Dixit},
  \citenamefont {Chakram}, \citenamefont {He}, \citenamefont {Agrawal},
  \citenamefont {Naik}, \citenamefont {Schuster},\ and\ \citenamefont
  {Chou}}]{dixit_searching_2021}%
  \BibitemOpen
  \bibfield  {author} {\bibinfo {author} {\bibfnamefont {A.~V.}\ \bibnamefont
  {Dixit}}, \bibinfo {author} {\bibfnamefont {S.}~\bibnamefont {Chakram}},
  \bibinfo {author} {\bibfnamefont {K.}~\bibnamefont {He}}, \bibinfo {author}
  {\bibfnamefont {A.}~\bibnamefont {Agrawal}}, \bibinfo {author} {\bibfnamefont
  {R.~K.}\ \bibnamefont {Naik}}, \bibinfo {author} {\bibfnamefont {D.~I.}\
  \bibnamefont {Schuster}}, \ and\ \bibinfo {author} {\bibfnamefont
  {A.}~\bibnamefont {Chou}},\ }\href {\doibase 10.1103/PhysRevLett.126.141302}
  {\ \textbf {\bibinfo {volume} {126}},\ \bibinfo {pages} {141302}},\ \Eprint
  {http://arxiv.org/abs/2008.12231 [hep-ex, physics:quant-ph]} {2008.12231
  [hep-ex, physics:quant-ph]} \BibitemShut {NoStop}%
\bibitem [{\citenamefont {Scigliuzzo}\ \emph {et~al.}(2020)\citenamefont
  {Scigliuzzo}, \citenamefont {Bengtsson}, \citenamefont {Besse}, \citenamefont
  {Wallraff}, \citenamefont {Delsing},\ and\ \citenamefont
  {Gasparinetti}}]{scigliuzzo_primary_2020}%
  \BibitemOpen
  \bibfield  {author} {\bibinfo {author} {\bibfnamefont {M.}~\bibnamefont
  {Scigliuzzo}}, \bibinfo {author} {\bibfnamefont {A.}~\bibnamefont
  {Bengtsson}}, \bibinfo {author} {\bibfnamefont {J.-C.}\ \bibnamefont
  {Besse}}, \bibinfo {author} {\bibfnamefont {A.}~\bibnamefont {Wallraff}},
  \bibinfo {author} {\bibfnamefont {P.}~\bibnamefont {Delsing}}, \ and\
  \bibinfo {author} {\bibfnamefont {S.}~\bibnamefont {Gasparinetti}},\ }\href
  {\doibase 10.1103/PhysRevX.10.041054} {\bibfield  {journal} {\bibinfo
  {journal} {Physical Review X}\ }\textbf {\bibinfo {volume} {10}},\ \bibinfo
  {pages} {041054} (\bibinfo {year} {2020})},\ \bibinfo {note} {publisher:
  American Physical Society}\BibitemShut {NoStop}%
\bibitem [{\citenamefont {Assouly}\ \emph {et~al.}(2023)\citenamefont
  {Assouly}, \citenamefont {Dassonneville}, \citenamefont {Peronnin},
  \citenamefont {Bienfait},\ and\ \citenamefont
  {Huard}}]{assouly_quantum_2023}%
  \BibitemOpen
  \bibfield  {author} {\bibinfo {author} {\bibfnamefont {R.}~\bibnamefont
  {Assouly}}, \bibinfo {author} {\bibfnamefont {R.}~\bibnamefont
  {Dassonneville}}, \bibinfo {author} {\bibfnamefont {T.}~\bibnamefont
  {Peronnin}}, \bibinfo {author} {\bibfnamefont {A.}~\bibnamefont {Bienfait}},
  \ and\ \bibinfo {author} {\bibfnamefont {B.}~\bibnamefont {Huard}},\ }\href
  {\doibase 10.1038/s41567-023-02113-4} {\bibfield  {journal} {\bibinfo
  {journal} {Nature Physics}\ }\textbf {\bibinfo {volume} {19}},\ \bibinfo
  {pages} {1418} (\bibinfo {year} {2023})},\ \bibinfo {note} {number: 10
  Publisher: Nature Publishing Group}\BibitemShut {NoStop}%
\bibitem [{\citenamefont {Raussendorf}\ \emph {et~al.}()\citenamefont
  {Raussendorf}, \citenamefont {Browne},\ and\ \citenamefont
  {Briegel}}]{raussendorf_measurement-based_2003}%
  \BibitemOpen
  \bibfield  {author} {\bibinfo {author} {\bibfnamefont {R.}~\bibnamefont
  {Raussendorf}}, \bibinfo {author} {\bibfnamefont {D.~E.}\ \bibnamefont
  {Browne}}, \ and\ \bibinfo {author} {\bibfnamefont {H.~J.}\ \bibnamefont
  {Briegel}},\ }\href {\doibase 10.1103/PhysRevA.68.022312} {\ \textbf
  {\bibinfo {volume} {68}},\ \bibinfo {pages} {022312}}\BibitemShut {NoStop}%
\bibitem [{\citenamefont {Briegel}\ \emph {et~al.}()\citenamefont {Briegel},
  \citenamefont {Browne}, \citenamefont {Dür}, \citenamefont {Raussendorf},\
  and\ \citenamefont {Van~den Nest}}]{briegel_measurement-based_2009}%
  \BibitemOpen
  \bibfield  {author} {\bibinfo {author} {\bibfnamefont {H.~J.}\ \bibnamefont
  {Briegel}}, \bibinfo {author} {\bibfnamefont {D.~E.}\ \bibnamefont {Browne}},
  \bibinfo {author} {\bibfnamefont {W.}~\bibnamefont {Dür}}, \bibinfo {author}
  {\bibfnamefont {R.}~\bibnamefont {Raussendorf}}, \ and\ \bibinfo {author}
  {\bibfnamefont {M.}~\bibnamefont {Van~den Nest}},\ }\href {\doibase
  10.1038/nphys1157} {\ \textbf {\bibinfo {volume} {5}},\ \bibinfo {pages}
  {19}},\ \bibinfo {note} {number: 1 Publisher: Nature Publishing
  Group}\BibitemShut {NoStop}%
\bibitem [{\citenamefont {Bartolucci}\ \emph {et~al.}()\citenamefont
  {Bartolucci}, \citenamefont {Birchall}, \citenamefont {Bombin}, \citenamefont
  {Cable}, \citenamefont {Dawson}, \citenamefont {Gimeno-Segovia},
  \citenamefont {Johnston}, \citenamefont {Kieling}, \citenamefont {Nickerson},
  \citenamefont {Pant}, \citenamefont {Pastawski}, \citenamefont {Rudolph},\
  and\ \citenamefont {Sparrow}}]{bartolucci_fusion-based_2021}%
  \BibitemOpen
  \bibfield  {author} {\bibinfo {author} {\bibfnamefont {S.}~\bibnamefont
  {Bartolucci}}, \bibinfo {author} {\bibfnamefont {P.}~\bibnamefont
  {Birchall}}, \bibinfo {author} {\bibfnamefont {H.}~\bibnamefont {Bombin}},
  \bibinfo {author} {\bibfnamefont {H.}~\bibnamefont {Cable}}, \bibinfo
  {author} {\bibfnamefont {C.}~\bibnamefont {Dawson}}, \bibinfo {author}
  {\bibfnamefont {M.}~\bibnamefont {Gimeno-Segovia}}, \bibinfo {author}
  {\bibfnamefont {E.}~\bibnamefont {Johnston}}, \bibinfo {author}
  {\bibfnamefont {K.}~\bibnamefont {Kieling}}, \bibinfo {author} {\bibfnamefont
  {N.}~\bibnamefont {Nickerson}}, \bibinfo {author} {\bibfnamefont
  {M.}~\bibnamefont {Pant}}, \bibinfo {author} {\bibfnamefont {F.}~\bibnamefont
  {Pastawski}}, \bibinfo {author} {\bibfnamefont {T.}~\bibnamefont {Rudolph}},
  \ and\ \bibinfo {author} {\bibfnamefont {C.}~\bibnamefont {Sparrow}},\ }\href
  {http://arxiv.org/abs/2101.09310} {\enquote {\bibinfo {title} {Fusion-based
  quantum computation},}\ }\Eprint {http://arxiv.org/abs/2101.09310 [quant-ph]}
  {2101.09310 [quant-ph]} \BibitemShut {NoStop}%
\bibitem [{\citenamefont {Narla}\ \emph {et~al.}(2016)\citenamefont {Narla},
  \citenamefont {Shankar}, \citenamefont {Hatridge}, \citenamefont {Leghtas},
  \citenamefont {Sliwa}, \citenamefont {Zalys-Geller}, \citenamefont
  {Mundhada}, \citenamefont {Pfaff}, \citenamefont {Frunzio}, \citenamefont
  {Schoelkopf},\ and\ \citenamefont {Devoret}}]{narla_robust_2016}%
  \BibitemOpen
  \bibfield  {author} {\bibinfo {author} {\bibfnamefont {A.}~\bibnamefont
  {Narla}}, \bibinfo {author} {\bibfnamefont {S.}~\bibnamefont {Shankar}},
  \bibinfo {author} {\bibfnamefont {M.}~\bibnamefont {Hatridge}}, \bibinfo
  {author} {\bibfnamefont {Z.}~\bibnamefont {Leghtas}}, \bibinfo {author}
  {\bibfnamefont {K.}~\bibnamefont {Sliwa}}, \bibinfo {author} {\bibfnamefont
  {E.}~\bibnamefont {Zalys-Geller}}, \bibinfo {author} {\bibfnamefont
  {S.}~\bibnamefont {Mundhada}}, \bibinfo {author} {\bibfnamefont
  {W.}~\bibnamefont {Pfaff}}, \bibinfo {author} {\bibfnamefont
  {L.}~\bibnamefont {Frunzio}}, \bibinfo {author} {\bibfnamefont
  {R.}~\bibnamefont {Schoelkopf}}, \ and\ \bibinfo {author} {\bibfnamefont
  {M.}~\bibnamefont {Devoret}},\ }\href {\doibase 10.1103/PhysRevX.6.031036}
  {\bibfield  {journal} {\bibinfo  {journal} {Physical Review X}\ }\textbf
  {\bibinfo {volume} {6}},\ \bibinfo {pages} {031036} (\bibinfo {year}
  {2016})}\BibitemShut {NoStop}%
\bibitem [{\citenamefont {Opremcak}\ \emph {et~al.}()\citenamefont {Opremcak},
  \citenamefont {Pechenezhskiy}, \citenamefont {Howington}, \citenamefont
  {Christensen}, \citenamefont {Beck}, \citenamefont {Leonard}, \citenamefont
  {Suttle}, \citenamefont {Wilen}, \citenamefont {Nesterov}, \citenamefont
  {Ribeill}, \citenamefont {Thorbeck}, \citenamefont {Schlenker}, \citenamefont
  {Vavilov}, \citenamefont {Plourde},\ and\ \citenamefont
  {{McDermott}}}]{opremcak_measurement_2018}%
  \BibitemOpen
  \bibfield  {author} {\bibinfo {author} {\bibfnamefont {A.}~\bibnamefont
  {Opremcak}}, \bibinfo {author} {\bibfnamefont {I.~V.}\ \bibnamefont
  {Pechenezhskiy}}, \bibinfo {author} {\bibfnamefont {C.}~\bibnamefont
  {Howington}}, \bibinfo {author} {\bibfnamefont {B.~G.}\ \bibnamefont
  {Christensen}}, \bibinfo {author} {\bibfnamefont {M.~A.}\ \bibnamefont
  {Beck}}, \bibinfo {author} {\bibfnamefont {E.}~\bibnamefont {Leonard}},
  \bibinfo {author} {\bibfnamefont {J.}~\bibnamefont {Suttle}}, \bibinfo
  {author} {\bibfnamefont {C.}~\bibnamefont {Wilen}}, \bibinfo {author}
  {\bibfnamefont {K.~N.}\ \bibnamefont {Nesterov}}, \bibinfo {author}
  {\bibfnamefont {G.~J.}\ \bibnamefont {Ribeill}}, \bibinfo {author}
  {\bibfnamefont {T.}~\bibnamefont {Thorbeck}}, \bibinfo {author}
  {\bibfnamefont {F.}~\bibnamefont {Schlenker}}, \bibinfo {author}
  {\bibfnamefont {M.~G.}\ \bibnamefont {Vavilov}}, \bibinfo {author}
  {\bibfnamefont {B.~L.~T.}\ \bibnamefont {Plourde}}, \ and\ \bibinfo {author}
  {\bibfnamefont {R.}~\bibnamefont {{McDermott}}},\ }\href {\doibase
  10.1126/science.aat4625} {\ \textbf {\bibinfo {volume} {361}},\ \bibinfo
  {pages} {1239}},\ \bibinfo {note} {publisher: American Association for the
  Advancement of Science}\BibitemShut {NoStop}%
\bibitem [{\citenamefont {Besse}\ \emph {et~al.}()\citenamefont {Besse},
  \citenamefont {Gasparinetti}, \citenamefont {Collodo}, \citenamefont
  {Walter}, \citenamefont {Remm}, \citenamefont {Krause}, \citenamefont
  {Eichler},\ and\ \citenamefont {Wallraff}}]{besse_parity_2020}%
  \BibitemOpen
  \bibfield  {author} {\bibinfo {author} {\bibfnamefont {J.-C.}\ \bibnamefont
  {Besse}}, \bibinfo {author} {\bibfnamefont {S.}~\bibnamefont {Gasparinetti}},
  \bibinfo {author} {\bibfnamefont {M.~C.}\ \bibnamefont {Collodo}}, \bibinfo
  {author} {\bibfnamefont {T.}~\bibnamefont {Walter}}, \bibinfo {author}
  {\bibfnamefont {A.}~\bibnamefont {Remm}}, \bibinfo {author} {\bibfnamefont
  {J.}~\bibnamefont {Krause}}, \bibinfo {author} {\bibfnamefont
  {C.}~\bibnamefont {Eichler}}, \ and\ \bibinfo {author} {\bibfnamefont
  {A.}~\bibnamefont {Wallraff}},\ }\href {\doibase 10.1103/PhysRevX.10.011046}
  {\ \textbf {\bibinfo {volume} {10}},\ \bibinfo {pages} {011046}}\BibitemShut
  {NoStop}%
\bibitem [{\citenamefont {Romero}\ \emph {et~al.}(2009)\citenamefont {Romero},
  \citenamefont {García-Ripoll},\ and\ \citenamefont
  {Solano}}]{romero_microwave_2009}%
  \BibitemOpen
  \bibfield  {author} {\bibinfo {author} {\bibfnamefont {G.}~\bibnamefont
  {Romero}}, \bibinfo {author} {\bibfnamefont {J.~J.}\ \bibnamefont
  {García-Ripoll}}, \ and\ \bibinfo {author} {\bibfnamefont {E.}~\bibnamefont
  {Solano}},\ }\href {\doibase 10.1103/PhysRevLett.102.173602} {\bibfield
  {journal} {\bibinfo  {journal} {Physical Review Letters}\ }\textbf {\bibinfo
  {volume} {102}},\ \bibinfo {pages} {173602} (\bibinfo {year} {2009})},\
  \bibinfo {note} {publisher: American Physical Society}\BibitemShut {NoStop}%
\bibitem [{\citenamefont {Helmer}\ \emph {et~al.}(2009)\citenamefont {Helmer},
  \citenamefont {Mariantoni}, \citenamefont {Solano},\ and\ \citenamefont
  {Marquardt}}]{helmer_quantum_2009}%
  \BibitemOpen
  \bibfield  {author} {\bibinfo {author} {\bibfnamefont {F.}~\bibnamefont
  {Helmer}}, \bibinfo {author} {\bibfnamefont {M.}~\bibnamefont {Mariantoni}},
  \bibinfo {author} {\bibfnamefont {E.}~\bibnamefont {Solano}}, \ and\ \bibinfo
  {author} {\bibfnamefont {F.}~\bibnamefont {Marquardt}},\ }\href {\doibase
  10.1103/PhysRevA.79.052115} {\bibfield  {journal} {\bibinfo  {journal}
  {Physical Review A}\ }\textbf {\bibinfo {volume} {79}},\ \bibinfo {pages}
  {052115} (\bibinfo {year} {2009})},\ \bibinfo {note} {publisher: American
  Physical Society}\BibitemShut {NoStop}%
\bibitem [{\citenamefont {Sathyamoorthy}\ \emph {et~al.}(2014)\citenamefont
  {Sathyamoorthy}, \citenamefont {Tornberg}, \citenamefont {Kockum},
  \citenamefont {Baragiola}, \citenamefont {Combes}, \citenamefont {Wilson},
  \citenamefont {Stace},\ and\ \citenamefont
  {Johansson}}]{sathyamoorthy_quantum_2014}%
  \BibitemOpen
  \bibfield  {author} {\bibinfo {author} {\bibfnamefont {S.~R.}\ \bibnamefont
  {Sathyamoorthy}}, \bibinfo {author} {\bibfnamefont {L.}~\bibnamefont
  {Tornberg}}, \bibinfo {author} {\bibfnamefont {A.~F.}\ \bibnamefont
  {Kockum}}, \bibinfo {author} {\bibfnamefont {B.~Q.}\ \bibnamefont
  {Baragiola}}, \bibinfo {author} {\bibfnamefont {J.}~\bibnamefont {Combes}},
  \bibinfo {author} {\bibfnamefont {C.}~\bibnamefont {Wilson}}, \bibinfo
  {author} {\bibfnamefont {T.~M.}\ \bibnamefont {Stace}}, \ and\ \bibinfo
  {author} {\bibfnamefont {G.}~\bibnamefont {Johansson}},\ }\href {\doibase
  10.1103/PhysRevLett.112.093601} {\bibfield  {journal} {\bibinfo  {journal}
  {Physical Review Letters}\ }\textbf {\bibinfo {volume} {112}},\ \bibinfo
  {pages} {093601} (\bibinfo {year} {2014})},\ \bibinfo {note} {publisher:
  American Physical Society}\BibitemShut {NoStop}%
\bibitem [{\citenamefont {Kyriienko}\ and\ \citenamefont
  {Sørensen}(2016)}]{kyriienko_continuous-wave_2016}%
  \BibitemOpen
  \bibfield  {author} {\bibinfo {author} {\bibfnamefont {O.}~\bibnamefont
  {Kyriienko}}\ and\ \bibinfo {author} {\bibfnamefont {A.~S.}\ \bibnamefont
  {Sørensen}},\ }\href {\doibase 10.1103/PhysRevLett.117.140503} {\bibfield
  {journal} {\bibinfo  {journal} {Physical Review Letters}\ }\textbf {\bibinfo
  {volume} {117}},\ \bibinfo {pages} {140503} (\bibinfo {year} {2016})},\
  \bibinfo {note} {publisher: American Physical Society}\BibitemShut {NoStop}%
\bibitem [{\citenamefont {Sathyamoorthy}\ \emph {et~al.}(2016)\citenamefont
  {Sathyamoorthy}, \citenamefont {Stace},\ and\ \citenamefont
  {Johansson}}]{sathyamoorthy_detecting_2016}%
  \BibitemOpen
  \bibfield  {author} {\bibinfo {author} {\bibfnamefont {S.~R.}\ \bibnamefont
  {Sathyamoorthy}}, \bibinfo {author} {\bibfnamefont {T.~M.}\ \bibnamefont
  {Stace}}, \ and\ \bibinfo {author} {\bibfnamefont {G.}~\bibnamefont
  {Johansson}},\ }\href {\doibase 10.1016/j.crhy.2016.07.010} {\bibfield
  {journal} {\bibinfo  {journal} {Comptes Rendus Physique}\ }\bibinfo {series}
  {Quantum microwaves / {Micro}-ondes quantiques},\ \textbf {\bibinfo {volume}
  {17}},\ \bibinfo {pages} {756} (\bibinfo {year} {2016})}\BibitemShut
  {NoStop}%
\bibitem [{\citenamefont {Gu}\ \emph {et~al.}(2017)\citenamefont {Gu},
  \citenamefont {Kockum}, \citenamefont {Miranowicz}, \citenamefont {Liu},\
  and\ \citenamefont {Nori}}]{gu_microwave_2017}%
  \BibitemOpen
  \bibfield  {author} {\bibinfo {author} {\bibfnamefont {X.}~\bibnamefont
  {Gu}}, \bibinfo {author} {\bibfnamefont {A.~F.}\ \bibnamefont {Kockum}},
  \bibinfo {author} {\bibfnamefont {A.}~\bibnamefont {Miranowicz}}, \bibinfo
  {author} {\bibfnamefont {Y.-x.}\ \bibnamefont {Liu}}, \ and\ \bibinfo
  {author} {\bibfnamefont {F.}~\bibnamefont {Nori}},\ }\href {\doibase
  10.1016/j.physrep.2017.10.002} {\bibfield  {journal} {\bibinfo  {journal}
  {Physics Reports}\ }\bibinfo {series} {Microwave photonics with
  superconducting quantum circuits},\ \textbf {\bibinfo {volume} {718-719}},\
  \bibinfo {pages} {1} (\bibinfo {year} {2017})}\BibitemShut {NoStop}%
\bibitem [{\citenamefont {Wong}\ and\ \citenamefont
  {Vavilov}(2017)}]{wong_quantum_2017}%
  \BibitemOpen
  \bibfield  {author} {\bibinfo {author} {\bibfnamefont {C.~H.}\ \bibnamefont
  {Wong}}\ and\ \bibinfo {author} {\bibfnamefont {M.~G.}\ \bibnamefont
  {Vavilov}},\ }\href {\doibase 10.1103/PhysRevA.95.012325} {\bibfield
  {journal} {\bibinfo  {journal} {Physical Review A}\ }\textbf {\bibinfo
  {volume} {95}},\ \bibinfo {pages} {012325} (\bibinfo {year} {2017})},\
  \bibinfo {note} {publisher: American Physical Society}\BibitemShut {NoStop}%
\bibitem [{\citenamefont {Royer}\ \emph {et~al.}(2018)\citenamefont {Royer},
  \citenamefont {Grimsmo}, \citenamefont {Choquette-Poitevin},\ and\
  \citenamefont {Blais}}]{royer_itinerant_2018}%
  \BibitemOpen
  \bibfield  {author} {\bibinfo {author} {\bibfnamefont {B.}~\bibnamefont
  {Royer}}, \bibinfo {author} {\bibfnamefont {A.~L.}\ \bibnamefont {Grimsmo}},
  \bibinfo {author} {\bibfnamefont {A.}~\bibnamefont {Choquette-Poitevin}}, \
  and\ \bibinfo {author} {\bibfnamefont {A.}~\bibnamefont {Blais}},\ }\href
  {\doibase 10.1103/PhysRevLett.120.203602} {\bibfield  {journal} {\bibinfo
  {journal} {Physical Review Letters}\ }\textbf {\bibinfo {volume} {120}},\
  \bibinfo {pages} {203602} (\bibinfo {year} {2018})},\ \bibinfo {note} {arXiv:
  1710.06040}\BibitemShut {NoStop}%
\bibitem [{\citenamefont {Grimsmo}\ \emph {et~al.}(2020)\citenamefont
  {Grimsmo}, \citenamefont {Royer}, \citenamefont {Kreikebaum}, \citenamefont
  {Ye}, \citenamefont {O'Brien}, \citenamefont {Siddiqi},\ and\ \citenamefont
  {Blais}}]{grimsmo_quantum_2020}%
  \BibitemOpen
  \bibfield  {author} {\bibinfo {author} {\bibfnamefont {A.~L.}\ \bibnamefont
  {Grimsmo}}, \bibinfo {author} {\bibfnamefont {B.}~\bibnamefont {Royer}},
  \bibinfo {author} {\bibfnamefont {J.~M.}\ \bibnamefont {Kreikebaum}},
  \bibinfo {author} {\bibfnamefont {Y.}~\bibnamefont {Ye}}, \bibinfo {author}
  {\bibfnamefont {K.}~\bibnamefont {O'Brien}}, \bibinfo {author} {\bibfnamefont
  {I.}~\bibnamefont {Siddiqi}}, \ and\ \bibinfo {author} {\bibfnamefont
  {A.}~\bibnamefont {Blais}},\ }\href {http://arxiv.org/abs/2005.06483}
  {\bibfield  {journal} {\bibinfo  {journal} {arXiv:2005.06483 [quant-ph]}\ }
  (\bibinfo {year} {2020})},\ \bibinfo {note} {arXiv: 2005.06483}\BibitemShut
  {NoStop}%
\bibitem [{\citenamefont {Chen}\ \emph {et~al.}(2011)\citenamefont {Chen},
  \citenamefont {Hover}, \citenamefont {Sendelbach}, \citenamefont {Maurer},
  \citenamefont {Merkel}, \citenamefont {Pritchett}, \citenamefont {Wilhelm},\
  and\ \citenamefont {McDermott}}]{chen_microwave_2011}%
  \BibitemOpen
  \bibfield  {author} {\bibinfo {author} {\bibfnamefont {Y.-F.}\ \bibnamefont
  {Chen}}, \bibinfo {author} {\bibfnamefont {D.}~\bibnamefont {Hover}},
  \bibinfo {author} {\bibfnamefont {S.}~\bibnamefont {Sendelbach}}, \bibinfo
  {author} {\bibfnamefont {L.}~\bibnamefont {Maurer}}, \bibinfo {author}
  {\bibfnamefont {S.~T.}\ \bibnamefont {Merkel}}, \bibinfo {author}
  {\bibfnamefont {E.~J.}\ \bibnamefont {Pritchett}}, \bibinfo {author}
  {\bibfnamefont {F.~K.}\ \bibnamefont {Wilhelm}}, \ and\ \bibinfo {author}
  {\bibfnamefont {R.}~\bibnamefont {McDermott}},\ }\href {\doibase
  10.1103/PhysRevLett.107.217401} {\bibfield  {journal} {\bibinfo  {journal}
  {Physical Review Letters}\ }\textbf {\bibinfo {volume} {107}},\ \bibinfo
  {pages} {217401} (\bibinfo {year} {2011})}\BibitemShut {NoStop}%
\bibitem [{\citenamefont {Koshino}\ \emph {et~al.}(2013)\citenamefont
  {Koshino}, \citenamefont {Inomata}, \citenamefont {Yamamoto},\ and\
  \citenamefont {Nakamura}}]{koshino_implementation_2013}%
  \BibitemOpen
  \bibfield  {author} {\bibinfo {author} {\bibfnamefont {K.}~\bibnamefont
  {Koshino}}, \bibinfo {author} {\bibfnamefont {K.}~\bibnamefont {Inomata}},
  \bibinfo {author} {\bibfnamefont {T.}~\bibnamefont {Yamamoto}}, \ and\
  \bibinfo {author} {\bibfnamefont {Y.}~\bibnamefont {Nakamura}},\ }\href
  {\doibase 10.1103/PhysRevLett.111.153601} {\bibfield  {journal} {\bibinfo
  {journal} {Physical Review Letters}\ }\textbf {\bibinfo {volume} {111}},\
  \bibinfo {pages} {153601} (\bibinfo {year} {2013})},\ \bibinfo {note}
  {publisher: American Physical Society}\BibitemShut {NoStop}%
\bibitem [{\citenamefont {Inomata}\ \emph {et~al.}(2016)\citenamefont
  {Inomata}, \citenamefont {Lin}, \citenamefont {Koshino}, \citenamefont
  {Oliver}, \citenamefont {Tsai}, \citenamefont {Yamamoto},\ and\ \citenamefont
  {Nakamura}}]{inomata_single_2016}%
  \BibitemOpen
  \bibfield  {author} {\bibinfo {author} {\bibfnamefont {K.}~\bibnamefont
  {Inomata}}, \bibinfo {author} {\bibfnamefont {Z.}~\bibnamefont {Lin}},
  \bibinfo {author} {\bibfnamefont {K.}~\bibnamefont {Koshino}}, \bibinfo
  {author} {\bibfnamefont {W.~D.}\ \bibnamefont {Oliver}}, \bibinfo {author}
  {\bibfnamefont {J.-S.}\ \bibnamefont {Tsai}}, \bibinfo {author}
  {\bibfnamefont {T.}~\bibnamefont {Yamamoto}}, \ and\ \bibinfo {author}
  {\bibfnamefont {Y.}~\bibnamefont {Nakamura}},\ }\href {\doibase
  10.1038/ncomms12303} {\bibfield  {journal} {\bibinfo  {journal} {Nature
  Communications}\ }\textbf {\bibinfo {volume} {7}},\ \bibinfo {pages} {12303}
  (\bibinfo {year} {2016})}\BibitemShut {NoStop}%
\bibitem [{\citenamefont {Besse}\ \emph {et~al.}(2018)\citenamefont {Besse},
  \citenamefont {Gasparinetti}, \citenamefont {Collodo}, \citenamefont
  {Walter}, \citenamefont {Kurpiers}, \citenamefont {Pechal}, \citenamefont
  {Eichler},\ and\ \citenamefont {Wallraff}}]{besse_single-shot_2018}%
  \BibitemOpen
  \bibfield  {author} {\bibinfo {author} {\bibfnamefont {J.-C.}\ \bibnamefont
  {Besse}}, \bibinfo {author} {\bibfnamefont {S.}~\bibnamefont {Gasparinetti}},
  \bibinfo {author} {\bibfnamefont {M.~C.}\ \bibnamefont {Collodo}}, \bibinfo
  {author} {\bibfnamefont {T.}~\bibnamefont {Walter}}, \bibinfo {author}
  {\bibfnamefont {P.}~\bibnamefont {Kurpiers}}, \bibinfo {author}
  {\bibfnamefont {M.}~\bibnamefont {Pechal}}, \bibinfo {author} {\bibfnamefont
  {C.}~\bibnamefont {Eichler}}, \ and\ \bibinfo {author} {\bibfnamefont
  {A.}~\bibnamefont {Wallraff}},\ }\href {\doibase 10.1103/PhysRevX.8.021003}
  {\bibfield  {journal} {\bibinfo  {journal} {Physical Review X}\ }\textbf
  {\bibinfo {volume} {8}},\ \bibinfo {pages} {021003} (\bibinfo {year}
  {2018})}\BibitemShut {NoStop}%
\bibitem [{\citenamefont {Kono}\ \emph {et~al.}(2018)\citenamefont {Kono},
  \citenamefont {Koshino}, \citenamefont {Tabuchi}, \citenamefont {Noguchi},\
  and\ \citenamefont {Nakamura}}]{kono_quantum_2018}%
  \BibitemOpen
  \bibfield  {author} {\bibinfo {author} {\bibfnamefont {S.}~\bibnamefont
  {Kono}}, \bibinfo {author} {\bibfnamefont {K.}~\bibnamefont {Koshino}},
  \bibinfo {author} {\bibfnamefont {Y.}~\bibnamefont {Tabuchi}}, \bibinfo
  {author} {\bibfnamefont {A.}~\bibnamefont {Noguchi}}, \ and\ \bibinfo
  {author} {\bibfnamefont {Y.}~\bibnamefont {Nakamura}},\ }\href {\doibase
  10.1038/s41567-018-0066-3} {\bibfield  {journal} {\bibinfo  {journal} {Nature
  Physics}\ }\textbf {\bibinfo {volume} {14}},\ \bibinfo {pages} {546}
  (\bibinfo {year} {2018})}\BibitemShut {NoStop}%
\bibitem [{\citenamefont {Lee}\ \emph {et~al.}()\citenamefont {Lee},
  \citenamefont {Efetov}, \citenamefont {Jung}, \citenamefont {Ranzani},
  \citenamefont {Walsh}, \citenamefont {Ohki}, \citenamefont {Taniguchi},
  \citenamefont {Watanabe}, \citenamefont {Kim}, \citenamefont {Englund},\ and\
  \citenamefont {Fong}}]{lee_graphene-based_2020}%
  \BibitemOpen
  \bibfield  {author} {\bibinfo {author} {\bibfnamefont {G.-H.}\ \bibnamefont
  {Lee}}, \bibinfo {author} {\bibfnamefont {D.~K.}\ \bibnamefont {Efetov}},
  \bibinfo {author} {\bibfnamefont {W.}~\bibnamefont {Jung}}, \bibinfo {author}
  {\bibfnamefont {L.}~\bibnamefont {Ranzani}}, \bibinfo {author} {\bibfnamefont
  {E.~D.}\ \bibnamefont {Walsh}}, \bibinfo {author} {\bibfnamefont {T.~A.}\
  \bibnamefont {Ohki}}, \bibinfo {author} {\bibfnamefont {T.}~\bibnamefont
  {Taniguchi}}, \bibinfo {author} {\bibfnamefont {K.}~\bibnamefont {Watanabe}},
  \bibinfo {author} {\bibfnamefont {P.}~\bibnamefont {Kim}}, \bibinfo {author}
  {\bibfnamefont {D.}~\bibnamefont {Englund}}, \ and\ \bibinfo {author}
  {\bibfnamefont {K.~C.}\ \bibnamefont {Fong}},\ }\href {\doibase
  10.1038/s41586-020-2752-4} {\ \textbf {\bibinfo {volume} {586}},\ \bibinfo
  {pages} {42}},\ \bibinfo {note} {number: 7827 Publisher: Nature Publishing
  Group}\BibitemShut {NoStop}%
\bibitem [{\citenamefont {Schuster}\ \emph {et~al.}(2007)\citenamefont
  {Schuster}, \citenamefont {Houck}, \citenamefont {Schreier}, \citenamefont
  {Wallraff}, \citenamefont {Gambetta}, \citenamefont {Blais}, \citenamefont
  {Frunzio}, \citenamefont {Majer}, \citenamefont {Johnson}, \citenamefont
  {Devoret}, \citenamefont {Girvin},\ and\ \citenamefont
  {Schoelkopf}}]{Schuster2007-zt}%
  \BibitemOpen
  \bibfield  {author} {\bibinfo {author} {\bibfnamefont {D.~I.}\ \bibnamefont
  {Schuster}}, \bibinfo {author} {\bibfnamefont {A.~A.}\ \bibnamefont {Houck}},
  \bibinfo {author} {\bibfnamefont {J.~A.}\ \bibnamefont {Schreier}}, \bibinfo
  {author} {\bibfnamefont {A.}~\bibnamefont {Wallraff}}, \bibinfo {author}
  {\bibfnamefont {J.~M.}\ \bibnamefont {Gambetta}}, \bibinfo {author}
  {\bibfnamefont {A.}~\bibnamefont {Blais}}, \bibinfo {author} {\bibfnamefont
  {L.}~\bibnamefont {Frunzio}}, \bibinfo {author} {\bibfnamefont
  {J.}~\bibnamefont {Majer}}, \bibinfo {author} {\bibfnamefont
  {B.}~\bibnamefont {Johnson}}, \bibinfo {author} {\bibfnamefont {M.~H.}\
  \bibnamefont {Devoret}}, \bibinfo {author} {\bibfnamefont {S.~M.}\
  \bibnamefont {Girvin}}, \ and\ \bibinfo {author} {\bibfnamefont {R.~J.}\
  \bibnamefont {Schoelkopf}},\ }\href@noop {} {\bibfield  {journal} {\bibinfo
  {journal} {Nature}\ }\textbf {\bibinfo {volume} {445}},\ \bibinfo {pages}
  {515} (\bibinfo {year} {2007})}\BibitemShut {NoStop}%
\bibitem [{\citenamefont {Gleyzes}\ \emph {et~al.}(2007)\citenamefont
  {Gleyzes}, \citenamefont {Kuhr}, \citenamefont {Guerlin}, \citenamefont
  {Bernu}, \citenamefont {Del{\'e}glise}, \citenamefont {Busk~Hoff},
  \citenamefont {Brune}, \citenamefont {Raimond},\ and\ \citenamefont
  {Haroche}}]{Gleyzes2007-xp}%
  \BibitemOpen
  \bibfield  {author} {\bibinfo {author} {\bibfnamefont {S.}~\bibnamefont
  {Gleyzes}}, \bibinfo {author} {\bibfnamefont {S.}~\bibnamefont {Kuhr}},
  \bibinfo {author} {\bibfnamefont {C.}~\bibnamefont {Guerlin}}, \bibinfo
  {author} {\bibfnamefont {J.}~\bibnamefont {Bernu}}, \bibinfo {author}
  {\bibfnamefont {S.}~\bibnamefont {Del{\'e}glise}}, \bibinfo {author}
  {\bibfnamefont {U.}~\bibnamefont {Busk~Hoff}}, \bibinfo {author}
  {\bibfnamefont {M.}~\bibnamefont {Brune}}, \bibinfo {author} {\bibfnamefont
  {J.-M.}\ \bibnamefont {Raimond}}, \ and\ \bibinfo {author} {\bibfnamefont
  {S.}~\bibnamefont {Haroche}},\ }\href@noop {} {\bibfield  {journal} {\bibinfo
   {journal} {Nature}\ }\textbf {\bibinfo {volume} {446}},\ \bibinfo {pages}
  {297} (\bibinfo {year} {2007})}\BibitemShut {NoStop}%
\bibitem [{\citenamefont {Lescanne}\ \emph {et~al.}(2020)\citenamefont
  {Lescanne}, \citenamefont {Deléglise}, \citenamefont {Albertinale},
  \citenamefont {Réglade}, \citenamefont {Capelle}, \citenamefont {Ivanov},
  \citenamefont {Jacqmin}, \citenamefont {Leghtas},\ and\ \citenamefont
  {Flurin}}]{lescanne_irreversible_2020}%
  \BibitemOpen
  \bibfield  {author} {\bibinfo {author} {\bibfnamefont {R.}~\bibnamefont
  {Lescanne}}, \bibinfo {author} {\bibfnamefont {S.}~\bibnamefont
  {Deléglise}}, \bibinfo {author} {\bibfnamefont {E.}~\bibnamefont
  {Albertinale}}, \bibinfo {author} {\bibfnamefont {U.}~\bibnamefont
  {Réglade}}, \bibinfo {author} {\bibfnamefont {T.}~\bibnamefont {Capelle}},
  \bibinfo {author} {\bibfnamefont {E.}~\bibnamefont {Ivanov}}, \bibinfo
  {author} {\bibfnamefont {T.}~\bibnamefont {Jacqmin}}, \bibinfo {author}
  {\bibfnamefont {Z.}~\bibnamefont {Leghtas}}, \ and\ \bibinfo {author}
  {\bibfnamefont {E.}~\bibnamefont {Flurin}},\ }\href {\doibase
  10.1103/PhysRevX.10.021038} {\bibfield  {journal} {\bibinfo  {journal}
  {Physical Review X}\ }\textbf {\bibinfo {volume} {10}},\ \bibinfo {pages}
  {021038} (\bibinfo {year} {2020})}\BibitemShut {NoStop}%
\bibitem [{\citenamefont {Sete}\ \emph {et~al.}(2015)\citenamefont {Sete},
  \citenamefont {Martinis},\ and\ \citenamefont
  {Korotkov}}]{korotkov_Purcell_filter_2015}%
  \BibitemOpen
  \bibfield  {author} {\bibinfo {author} {\bibfnamefont {E.~A.}\ \bibnamefont
  {Sete}}, \bibinfo {author} {\bibfnamefont {J.~M.}\ \bibnamefont {Martinis}},
  \ and\ \bibinfo {author} {\bibfnamefont {A.~N.}\ \bibnamefont {Korotkov}},\
  }\href {\doibase 10.1103/PhysRevA.92.012325} {\bibfield  {journal} {\bibinfo
  {journal} {Phys. Rev. A}\ }\textbf {\bibinfo {volume} {92}},\ \bibinfo
  {pages} {012325} (\bibinfo {year} {2015})}\BibitemShut {NoStop}%
\bibitem [{\citenamefont {Serniak}\ \emph {et~al.}(2018)\citenamefont
  {Serniak}, \citenamefont {Hays}, \citenamefont {de~Lange}, \citenamefont
  {Diamond}, \citenamefont {Shankar}, \citenamefont {Burkhart}, \citenamefont
  {Frunzio}, \citenamefont {Houzet},\ and\ \citenamefont
  {Devoret}}]{serniak_hot_2018}%
  \BibitemOpen
  \bibfield  {author} {\bibinfo {author} {\bibfnamefont {K.}~\bibnamefont
  {Serniak}}, \bibinfo {author} {\bibfnamefont {M.}~\bibnamefont {Hays}},
  \bibinfo {author} {\bibfnamefont {G.}~\bibnamefont {de~Lange}}, \bibinfo
  {author} {\bibfnamefont {S.}~\bibnamefont {Diamond}}, \bibinfo {author}
  {\bibfnamefont {S.}~\bibnamefont {Shankar}}, \bibinfo {author} {\bibfnamefont
  {L.~D.}\ \bibnamefont {Burkhart}}, \bibinfo {author} {\bibfnamefont
  {L.}~\bibnamefont {Frunzio}}, \bibinfo {author} {\bibfnamefont
  {M.}~\bibnamefont {Houzet}}, \ and\ \bibinfo {author} {\bibfnamefont {M.~H.}\
  \bibnamefont {Devoret}},\ }\href {\doibase 10.1103/PhysRevLett.121.157701}
  {\bibfield  {journal} {\bibinfo  {journal} {Physical Review Letters}\
  }\textbf {\bibinfo {volume} {121}},\ \bibinfo {pages} {157701} (\bibinfo
  {year} {2018})},\ \bibinfo {note} {arXiv: 1803.00476}\BibitemShut {NoStop}%
\bibitem [{\citenamefont {Serniak}\ \emph {et~al.}(2019)\citenamefont
  {Serniak}, \citenamefont {Diamond}, \citenamefont {Hays}, \citenamefont
  {Fatemi}, \citenamefont {Shankar}, \citenamefont {Frunzio}, \citenamefont
  {Schoelkopf},\ and\ \citenamefont {Devoret}}]{serniak_direct_2019}%
  \BibitemOpen
  \bibfield  {author} {\bibinfo {author} {\bibfnamefont {K.}~\bibnamefont
  {Serniak}}, \bibinfo {author} {\bibfnamefont {S.}~\bibnamefont {Diamond}},
  \bibinfo {author} {\bibfnamefont {M.}~\bibnamefont {Hays}}, \bibinfo {author}
  {\bibfnamefont {V.}~\bibnamefont {Fatemi}}, \bibinfo {author} {\bibfnamefont
  {S.}~\bibnamefont {Shankar}}, \bibinfo {author} {\bibfnamefont
  {L.}~\bibnamefont {Frunzio}}, \bibinfo {author} {\bibfnamefont
  {R.}~\bibnamefont {Schoelkopf}}, \ and\ \bibinfo {author} {\bibfnamefont
  {M.}~\bibnamefont {Devoret}},\ }\href {\doibase
  10.1103/PhysRevApplied.12.014052} {\bibfield  {journal} {\bibinfo  {journal}
  {Phys. Rev. Appl.}\ }\textbf {\bibinfo {volume} {12}},\ \bibinfo {pages}
  {014052} (\bibinfo {year} {2019})}\BibitemShut {NoStop}%
\bibitem [{\citenamefont {Jin}\ \emph {et~al.}(2015)\citenamefont {Jin},
  \citenamefont {Kamal}, \citenamefont {Sears}, \citenamefont {Gudmundsen},
  \citenamefont {Hover}, \citenamefont {Miloshi}, \citenamefont {Slattery},
  \citenamefont {Yan}, \citenamefont {Yoder}, \citenamefont {Orlando},
  \citenamefont {Gustavsson},\ and\ \citenamefont {Oliver}}]{Jin_Thermal_2015}%
  \BibitemOpen
  \bibfield  {author} {\bibinfo {author} {\bibfnamefont {X.~Y.}\ \bibnamefont
  {Jin}}, \bibinfo {author} {\bibfnamefont {A.}~\bibnamefont {Kamal}}, \bibinfo
  {author} {\bibfnamefont {A.~P.}\ \bibnamefont {Sears}}, \bibinfo {author}
  {\bibfnamefont {T.}~\bibnamefont {Gudmundsen}}, \bibinfo {author}
  {\bibfnamefont {D.}~\bibnamefont {Hover}}, \bibinfo {author} {\bibfnamefont
  {J.}~\bibnamefont {Miloshi}}, \bibinfo {author} {\bibfnamefont
  {R.}~\bibnamefont {Slattery}}, \bibinfo {author} {\bibfnamefont
  {F.}~\bibnamefont {Yan}}, \bibinfo {author} {\bibfnamefont {J.}~\bibnamefont
  {Yoder}}, \bibinfo {author} {\bibfnamefont {T.~P.}\ \bibnamefont {Orlando}},
  \bibinfo {author} {\bibfnamefont {S.}~\bibnamefont {Gustavsson}}, \ and\
  \bibinfo {author} {\bibfnamefont {W.~D.}\ \bibnamefont {Oliver}},\ }\href
  {\doibase 10.1103/PhysRevLett.114.240501} {\bibfield  {journal} {\bibinfo
  {journal} {Phys. Rev. Lett.}\ }\textbf {\bibinfo {volume} {114}},\ \bibinfo
  {pages} {240501} (\bibinfo {year} {2015})}\BibitemShut {NoStop}%
\bibitem [{\citenamefont {Connolly}\ \emph {et~al.}(2023)\citenamefont
  {Connolly}, \citenamefont {Kurilovich}, \citenamefont {Diamond},
  \citenamefont {Nho}, \citenamefont {Bøttcher}, \citenamefont {Glazman},
  \citenamefont {Fatemi},\ and\ \citenamefont
  {Devoret}}]{connolly_coexistence_2023}%
  \BibitemOpen
  \bibfield  {author} {\bibinfo {author} {\bibfnamefont {T.}~\bibnamefont
  {Connolly}}, \bibinfo {author} {\bibfnamefont {P.~D.}\ \bibnamefont
  {Kurilovich}}, \bibinfo {author} {\bibfnamefont {S.}~\bibnamefont {Diamond}},
  \bibinfo {author} {\bibfnamefont {H.}~\bibnamefont {Nho}}, \bibinfo {author}
  {\bibfnamefont {C.~G.~L.}\ \bibnamefont {Bøttcher}}, \bibinfo {author}
  {\bibfnamefont {L.~I.}\ \bibnamefont {Glazman}}, \bibinfo {author}
  {\bibfnamefont {V.}~\bibnamefont {Fatemi}}, \ and\ \bibinfo {author}
  {\bibfnamefont {M.~H.}\ \bibnamefont {Devoret}},\ }\href@noop {} {\
  (\bibinfo {year} {2023})},\ \Eprint {http://arxiv.org/abs/2302.12330}
  {arXiv:2302.12330 [quant-ph]} \BibitemShut {NoStop}%
\bibitem [{\citenamefont {Gambetta}\ \emph {et~al.}(2006)\citenamefont
  {Gambetta}, \citenamefont {Blais}, \citenamefont {Schuster}, \citenamefont
  {Wallraff}, \citenamefont {Frunzio}, \citenamefont {Majer}, \citenamefont
  {Devoret}, \citenamefont {Girvin},\ and\ \citenamefont
  {Schoelkopf}}]{gambetta_qubit-photon_2006}%
  \BibitemOpen
  \bibfield  {author} {\bibinfo {author} {\bibfnamefont {J.}~\bibnamefont
  {Gambetta}}, \bibinfo {author} {\bibfnamefont {A.}~\bibnamefont {Blais}},
  \bibinfo {author} {\bibfnamefont {D.~I.}\ \bibnamefont {Schuster}}, \bibinfo
  {author} {\bibfnamefont {A.}~\bibnamefont {Wallraff}}, \bibinfo {author}
  {\bibfnamefont {L.}~\bibnamefont {Frunzio}}, \bibinfo {author} {\bibfnamefont
  {J.}~\bibnamefont {Majer}}, \bibinfo {author} {\bibfnamefont {M.~H.}\
  \bibnamefont {Devoret}}, \bibinfo {author} {\bibfnamefont {S.~M.}\
  \bibnamefont {Girvin}}, \ and\ \bibinfo {author} {\bibfnamefont {R.~J.}\
  \bibnamefont {Schoelkopf}},\ }\href {\doibase 10.1103/PhysRevA.74.042318}
  {\bibfield  {journal} {\bibinfo  {journal} {Physical Review A}\ }\textbf
  {\bibinfo {volume} {74}},\ \bibinfo {pages} {042318} (\bibinfo {year}
  {2006})}\BibitemShut {NoStop}%
\bibitem [{\citenamefont {Balembois}(2023)}]{balembois_magnetic_2023}%
  \BibitemOpen
  \bibfield  {author} {\bibinfo {author} {\bibfnamefont {L.}~\bibnamefont
  {Balembois}},\ }\emph {\bibinfo {title} {Magnetic resonance of a single
  electron spin and its magnetic environment by photon counting}},\ \href@noop
  {} {\bibinfo {type} {thesis}},\ \bibinfo  {school} {Paris Saclay} (\bibinfo
  {year} {2023})\BibitemShut {NoStop}%
\bibitem [{\citenamefont {Place}\ \emph {et~al.}()\citenamefont {Place},
  \citenamefont {Rodgers}, \citenamefont {Mundada}, \citenamefont {Smitham},
  \citenamefont {Fitzpatrick}, \citenamefont {Leng}, \citenamefont {Premkumar},
  \citenamefont {Bryon}, \citenamefont {Vrajitoarea}, \citenamefont {Sussman},
  \citenamefont {Cheng}, \citenamefont {Madhavan}, \citenamefont {Babla},
  \citenamefont {Le}, \citenamefont {Gang}, \citenamefont {Jäck},
  \citenamefont {Gyenis}, \citenamefont {Yao}, \citenamefont {Cava},
  \citenamefont {de~Leon},\ and\ \citenamefont {Houck}}]{place_new_2021}%
  \BibitemOpen
  \bibfield  {author} {\bibinfo {author} {\bibfnamefont {A.~P.~M.}\
  \bibnamefont {Place}}, \bibinfo {author} {\bibfnamefont {L.~V.~H.}\
  \bibnamefont {Rodgers}}, \bibinfo {author} {\bibfnamefont {P.}~\bibnamefont
  {Mundada}}, \bibinfo {author} {\bibfnamefont {B.~M.}\ \bibnamefont
  {Smitham}}, \bibinfo {author} {\bibfnamefont {M.}~\bibnamefont
  {Fitzpatrick}}, \bibinfo {author} {\bibfnamefont {Z.}~\bibnamefont {Leng}},
  \bibinfo {author} {\bibfnamefont {A.}~\bibnamefont {Premkumar}}, \bibinfo
  {author} {\bibfnamefont {J.}~\bibnamefont {Bryon}}, \bibinfo {author}
  {\bibfnamefont {A.}~\bibnamefont {Vrajitoarea}}, \bibinfo {author}
  {\bibfnamefont {S.}~\bibnamefont {Sussman}}, \bibinfo {author} {\bibfnamefont
  {G.}~\bibnamefont {Cheng}}, \bibinfo {author} {\bibfnamefont
  {T.}~\bibnamefont {Madhavan}}, \bibinfo {author} {\bibfnamefont {H.~K.}\
  \bibnamefont {Babla}}, \bibinfo {author} {\bibfnamefont {X.~H.}\ \bibnamefont
  {Le}}, \bibinfo {author} {\bibfnamefont {Y.}~\bibnamefont {Gang}}, \bibinfo
  {author} {\bibfnamefont {B.}~\bibnamefont {Jäck}}, \bibinfo {author}
  {\bibfnamefont {A.}~\bibnamefont {Gyenis}}, \bibinfo {author} {\bibfnamefont
  {N.}~\bibnamefont {Yao}}, \bibinfo {author} {\bibfnamefont {R.~J.}\
  \bibnamefont {Cava}}, \bibinfo {author} {\bibfnamefont {N.~P.}\ \bibnamefont
  {de~Leon}}, \ and\ \bibinfo {author} {\bibfnamefont {A.~A.}\ \bibnamefont
  {Houck}},\ }\href {\doibase 10.1038/s41467-021-22030-5} {\ \textbf {\bibinfo
  {volume} {12}},\ \bibinfo {pages} {1779}},\ \bibinfo {note} {number: 1
  Publisher: Nature Publishing Group}\BibitemShut {NoStop}%
\bibitem [{\citenamefont {Wang}\ \emph {et~al.}()\citenamefont {Wang},
  \citenamefont {Li}, \citenamefont {Xu}, \citenamefont {Li}, \citenamefont
  {Wang}, \citenamefont {Yang}, \citenamefont {Mi}, \citenamefont {Liang},
  \citenamefont {Su}, \citenamefont {Yang}, \citenamefont {Wang}, \citenamefont
  {Wang}, \citenamefont {Li}, \citenamefont {Chen}, \citenamefont {Li},
  \citenamefont {Linghu}, \citenamefont {Han}, \citenamefont {Zhang},
  \citenamefont {Feng}, \citenamefont {Song}, \citenamefont {Ma}, \citenamefont
  {Zhang}, \citenamefont {Wang}, \citenamefont {Zhao}, \citenamefont {Liu},
  \citenamefont {Xue}, \citenamefont {Jin},\ and\ \citenamefont
  {Yu}}]{wang_towards_2022}%
  \BibitemOpen
  \bibfield  {author} {\bibinfo {author} {\bibfnamefont {C.}~\bibnamefont
  {Wang}}, \bibinfo {author} {\bibfnamefont {X.}~\bibnamefont {Li}}, \bibinfo
  {author} {\bibfnamefont {H.}~\bibnamefont {Xu}}, \bibinfo {author}
  {\bibfnamefont {Z.}~\bibnamefont {Li}}, \bibinfo {author} {\bibfnamefont
  {J.}~\bibnamefont {Wang}}, \bibinfo {author} {\bibfnamefont {Z.}~\bibnamefont
  {Yang}}, \bibinfo {author} {\bibfnamefont {Z.}~\bibnamefont {Mi}}, \bibinfo
  {author} {\bibfnamefont {X.}~\bibnamefont {Liang}}, \bibinfo {author}
  {\bibfnamefont {T.}~\bibnamefont {Su}}, \bibinfo {author} {\bibfnamefont
  {C.}~\bibnamefont {Yang}}, \bibinfo {author} {\bibfnamefont {G.}~\bibnamefont
  {Wang}}, \bibinfo {author} {\bibfnamefont {W.}~\bibnamefont {Wang}}, \bibinfo
  {author} {\bibfnamefont {Y.}~\bibnamefont {Li}}, \bibinfo {author}
  {\bibfnamefont {M.}~\bibnamefont {Chen}}, \bibinfo {author} {\bibfnamefont
  {C.}~\bibnamefont {Li}}, \bibinfo {author} {\bibfnamefont {K.}~\bibnamefont
  {Linghu}}, \bibinfo {author} {\bibfnamefont {J.}~\bibnamefont {Han}},
  \bibinfo {author} {\bibfnamefont {Y.}~\bibnamefont {Zhang}}, \bibinfo
  {author} {\bibfnamefont {Y.}~\bibnamefont {Feng}}, \bibinfo {author}
  {\bibfnamefont {Y.}~\bibnamefont {Song}}, \bibinfo {author} {\bibfnamefont
  {T.}~\bibnamefont {Ma}}, \bibinfo {author} {\bibfnamefont {J.}~\bibnamefont
  {Zhang}}, \bibinfo {author} {\bibfnamefont {R.}~\bibnamefont {Wang}},
  \bibinfo {author} {\bibfnamefont {P.}~\bibnamefont {Zhao}}, \bibinfo {author}
  {\bibfnamefont {W.}~\bibnamefont {Liu}}, \bibinfo {author} {\bibfnamefont
  {G.}~\bibnamefont {Xue}}, \bibinfo {author} {\bibfnamefont {Y.}~\bibnamefont
  {Jin}}, \ and\ \bibinfo {author} {\bibfnamefont {H.}~\bibnamefont {Yu}},\
  }\href {\doibase 10.1038/s41534-021-00510-2} {\ \textbf {\bibinfo {volume}
  {8}},\ \bibinfo {pages} {1}},\ \bibinfo {note} {number: 1 Publisher: Nature
  Publishing Group}\BibitemShut {NoStop}%
\end{thebibliography}%

\end{document}